\RequirePackage{fix-cm}
\documentclass[twocolumn,epjc3]{svjour3}  
\usepackage[utf8]{inputenc}
\usepackage[T1]{fontenc}
\smartqed  
\RequirePackage{graphicx}
\RequirePackage[numbers,sort&compress]{natbib}
\RequirePackage[colorlinks,citecolor=blue,urlcolor=blue,linkcolor=blue]{hyperref}

\RequirePackage{amsmath}
\RequirePackage{amssymb}
\RequirePackage[absolute]{textpos}
\RequirePackage{physics}
\RequirePackage{subfigure}
\RequirePackage{xspace}
\RequirePackage{tikz}
\tikzstyle{arrow} = [draw, -latex, thick]

\usepackage[colorinlistoftodos]{todonotes}

\journalname{Eur. Phys. J. C}

\hypersetup{
  pdftitle    = {Investigating multiple solutions to boundary value problems in constrained minimal and non-minimal SUSY models},
  pdfauthor   = {Daniel Meuser, Alexander Voigt},
  pdfkeywords = {MSSM, CMSSM, Supersymmetry, boundary value problem}
}

\newcommand{\eqn}[1]{Eq.\,(\ref{#1})}
\newcommand{\eqs}[1]{Eqs.\,(\ref{#1})}
\newcommand{\fig}[1]{Fig.\,\ref{#1}}
\newcommand{\figs}[1]{Figs.\,\ref{#1}}
\newcommand{\tab}[1]{Table\,\ref{#1}}

\newcommand{\sct}[1]{Sect.\,\ref{#1}}

\newcommand{\citere}[1]{Ref.~\cite{#1}}
\newcommand{\citeres}[1]{Refs.~\cite{#1}}

\DeclareMathOperator{\sign}{sign}
\newcommand{\GeV}{\,\ensuremath{\text{GeV}}}
\newcommand{\TeV}{\,\ensuremath{\text{TeV}}\xspace}
\newcommand{\gf}{G_F}
\newcommand{\mz}{M_Z}
\newcommand{\mw}{M_W}
\newcommand{\ms}{M_S}
\newcommand{\mx}{M_X}

\newcommand{\kallen}{Källén\xspace}
\newcommand{\mueff}{\mu_\text{eff}}
\newcommand{\FS}{\texttt{FlexibleSUSY}\xspace}

\newcommand{\DR}{\ensuremath{\overline{\text{DR}}'}\xspace}

\newcommand{\bigwhitestar}{\ensuremath{\bigstar\!\!\!\!\textcolor{white}{\star}}}

\DeclareFontFamily{\encodingdefault}{\ttdefault}{\hyphenchar\font=`\-}

\begin{document}

\begin{textblock*}{10em}(0.98\textwidth,2em)
\raggedleft\noindent\footnotesize
DESY 19-129 \\
TTK--19--28
\end{textblock*}

\title{Investigating multiple solutions to boundary value problems in
  constrained minimal and non-minimal SUSY models\thanksref{t1}}
\author{Daniel Meuser\thanksref{e1,addr1} \and Alexander
  Voigt\thanksref{e2,addr2}}

\thankstext{t1}{This research was supported by the German DFG Research
  Unit \textit{New Physics at the LHC} (FOR 2239) and the German DFG
  Excellence Strategy \textit{Quantum Universe} (EXC 2121).}
\thankstext{e1}{e-mail: daniel.meuser@desy.de}
\thankstext{e2}{e-mail: alexander.voigt@physik.rwth-aachen.de}

\institute{Deutsches Elektronen-Synchrotron DESY, Notkestra{\ss}e 85, 22607 Hamburg, Germany \label{addr1}
           \and
           Institute for Theoretical Particle Physics and Cosmology, RWTH Aachen University, 52074 Aachen, Germany \label{addr2}%
}

\date{Received: date / Accepted: date}

\maketitle

\begin{abstract}
  We investigate the physical origins of multiple solutions to
  boundary value problems in the fully constrained MSSM and NMSSM.  We
  derive mathematical criteria that formulate circumstances under
  which multiple solutions can appear.  Finally, we study the validity
  of the exclusion of the CMSSM in the presence of multiple solutions.
\end{abstract}

\section{Introduction}
\label{intro}

After the discovery of the Higgs boson with a mass of
$M_h = (125.10 \pm 0.14)\GeV$
\cite{Aad:2012tfa,Chatrchyan:2012xdj,Aad:2015zhl} and the
non-discovery of supersymmetric (SUSY) particles at the LHC, it becomes
clearer that pure weak-scale supersymmetry may not be realized in
nature.  Although the general Minimal Supersymmetric Standard Model
(MSSM) may be difficult to fully exclude, the constrained MSSM (CMSSM),
which is inspired by minimal supergravity,
has been excluded at more than $90\%$ confidence level
\cite{Bechtle:2015nua,Athron:2017qdc}.
One reason for the exclusion is that in the CMSSM, by construction, all
sfermion masses are of the same order.  In this case, however,
observables such as the Higgs boson mass, Dark Matter and the
anomalous magnetic moment of the muon cannot be explained
simultaneously by the MSSM, because $M_h \approx 125\GeV$ requires
multi-TeV stops, while the other observables prefer sub-TeV sleptons.

However, in \citeres{Allanach:2013yua,Allanach:2013cda,Allanach:2014sea}
it was discovered that there may be multiple MSSM parameter sets which
fulfill the same CMSSM boundary conditions.  The mathematical reason for this
phenomenon is that the CMSSM is formulated as a boundary value
problem (BVP), where the running MSSM \DR\ parameters are
fixed by input values at different renormalization scales.
The parameters at the
different scales are connected via a set of differential equations,
the so-called renormalization group equations (RGEs).  Formally, such
a BVP may have no, one, or multiple solutions for the MSSM parameters.

In order to make a
statement about the validity of the CMSSM, all possible solutions
to the BVP must be studied.  However, the BVP solving algorithm used
in the global fitting analyses of
\citeres{Bechtle:2015nua,Athron:2017qdc} can at most find one solution
and may miss further ones.  This raises the question whether the CMSSM
is still excluded in the presence of multiple solutions of the BVP.

In the present paper we systematically study the physical
origin of the multiple solutions in the CMSSM.  In doing this, we go
beyond the scope of \citeres{Allanach:2013yua,Allanach:2013cda} and
derive mathematical criteria that formulate circumstances under which
multiple solutions can appear.
In addition we study the influence of the chosen low-energy
observables that fix the electroweak gauge couplings, which appear to
play in important role for the occurrence and the number of multiple
solutions.
Next, we apply our newly gained insights to the results presented in
\citere{Bechtle:2015nua} and investigate the influence of multiple
solutions on the global fit performed therein.
Finally, we extend our analysis to the fully constrained Next-to-Minimal 
Supersymmetric Standard Model (CNMSSM) and demonstrate that
multiple solutions can also occur in constrained non-minimal SUSY
models. 

\section{Boundary value problems}

\subsection{CMSSM boundary conditions}
\label{sec:CMSSM_BPV}

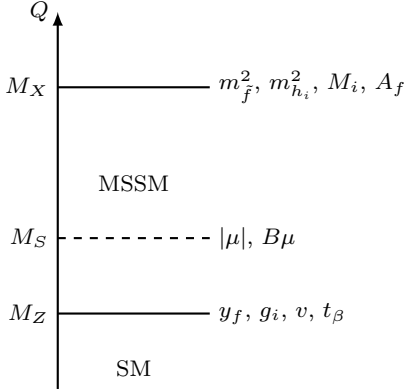
\begin{figure}[t]
  \centering
  \begin{tikzpicture}
    \path[arrow] (0,0) -- (0,5) node[left]{$Q$};
    \draw[thick] (0,4) node[left]{$\mx$} -- (2,4) node[right]{$m^2_{\tilde{f}}$, $m^2_{h_i}$, $M_i$, $A_f$};
    \draw[thick,dashed] (0,2) node[left]{$\ms$} -- (2,2) node[right]{$|\mu|$, $B\mu$};
    \draw[thick] (0,1) node[left]{$\mz$} -- node[above = 1.5cm]{MSSM} node[below = 0.5cm]{SM} (2,1) node[right]{$y_f$, $g_i$, $v$, $t_\beta$};
  \end{tikzpicture}
  \caption{CMSSM boundary value problem.}
  \label{fig:CMSSM_BVP}
\end{figure}
The CMSSM is formulated as a BVP, where the \DR parameters are fixed
at three different scales, see \fig{fig:CMSSM_BVP}.
At the electroweak scale $Q = \mz$, the \DR gauge and Yukawa couplings
$g_{i}$ ($i=1,2,3$) and $y_f$ as well as the SM-like Higgs
vacuum expectation value (VEV) $v=\sqrt{v_u^2 + v_d^2}$ are determined
from known Standard Model observables and the ratio $t_\beta = v_u/v_d$.
At the gauge coupling unification scale (GUT scale) $Q = \mx$, defined
by $g_1(\mx) = g_2(\mx)$, the soft-breaking sfermion and Higgs mass
parameters $m^2_{\tilde{f}}$ and $m^2_{h_{u,d}}$
($\tilde{f}=\tilde{q},\tilde{u},\tilde{d},\tilde{l},\tilde{e}$), the
gaugino mass parameters $M_i$ $(i=1,2,3)$ as well as the trilinear
couplings $A_f$ ($f=u,d,e$) are unified to
\begin{equation}
\begin{aligned}
  &(m^2_{\tilde{f}})_{ij} = m_0^2 \delta_{ij}, &&m^2_{h_u} = m^2_{h_d} = m_0^2, \\
  &M_i = M_{1/2}, &&(A_f)_{ij} = A_0 \delta_{ij}.
\end{aligned}
\label{eq:CMSSM_GUT_BC}
\end{equation}
At the SUSY scale
$Q = \ms \equiv \sqrt{m_{\tilde{t}_1}m_{\tilde{t}_2}}$, where
$m_{\tilde{t}_{i}}$ denotes the $i$-th \DR stop mass, the parameters
$|\mu|$ and $B\mu$ are fixed by the two electroweak symmetry breaking
(EWSB) equations.
This leaves the following five free parameters of the CMSSM:
\begin{equation}
  t_\beta(\mz), \, m_0^2, \, M_{1/2}, \, A_0, \, \sign(\mu).
\end{equation}
\subsection{CNMSSM boundary conditions}
\label{sec:CNMSSM_BVP}

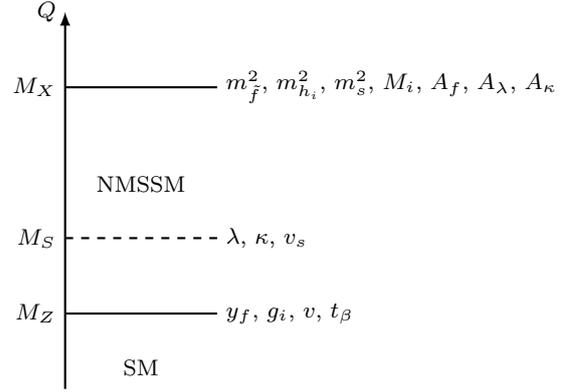
\begin{figure}[t]
  \centering
  \begin{tikzpicture}
    \path[arrow] (0,0) -- (0,5) node[left]{$Q$};
    \draw[thick] (0,4) node[left]{$\mx$} -- (2,4) node[right]{$m^2_{\tilde{f}}$, $m^2_{h_i}$, $m^2_s$, $M_i$, $A_f$, $A_\lambda$, $A_\kappa$};
    \draw[thick,dashed] (0,2) node[left]{$\ms$} -- (2,2) node[right]{$\lambda$, $\kappa$, $v_s$};
    \draw[thick] (0,1) node[left]{$\mz$} -- node[above = 1.5cm]{NMSSM} node[below = 0.5cm]{SM} (2,1) node[right]{$y_f$, $g_i$, $v$, $t_\beta$};
  \end{tikzpicture}
  \caption{CNMSSM boundary value problem.}
  \label{fig:CNSSM_BVP}
\end{figure}
In the $Z_3$ symmetric NMSSM the parameters $\mu$ and $B\mu$ are
absent and are replaced by the new parameters $\lambda$, $\kappa$,
$v_s$, $m_s^2$, $A_\lambda$ and $A_\kappa$ \cite{Ellwanger:2009dp}.
In the following study we consider the fully constrained $Z_3$
symmetric NMSSM (CNMSSM)
\cite{Ellwanger:1993xa,Ellwanger:1995ru,Ellwanger:1996gw,Djouadi:2008yj,Djouadi:2008uj},
where all soft-breaking parameters are unified at the GUT scale as in
\eqs{eq:CMSSM_GUT_BC} with the additional constraints
\begin{equation}
  m_s^2 = m_0^2, \quad A_\lambda = A_\kappa = A_0,
\end{equation}
see \fig{fig:CNSSM_BVP}.
This leaves seven free parameters $t_\beta$, $m_0^2$, $M_{1/2}$, $A_0$,
$\lambda$, $\kappa$, and $v_s$, of which three are fixed
by the NMSSM EWSB equations.  Since $\lambda$ and $\kappa$ are
dimensionless parameters that enter all NMSSM $\beta$ functions at
sufficiently high order, we take
them as input at $Q=\ms$ and fix the parameters $m_0^2$, $M_{1/2}$ and $A_0$ by
the three EWSB equations via the semi-analytic approach, see below.
As a result, we are left with the following four free CNMSSM
parameters:
\begin{equation}
  t_\beta(\mz), \, \lambda(\ms), \, \kappa(\ms), \, v_s(\ms) .
\end{equation}
In the following we trade $v_s$ for $\mueff = \lambda v_s/\sqrt{2}$.

\subsection{Matching to low-energy observables}

For the study of multiple solutions in the C(N)MSSM, the calculation of
the \DR electroweak gauge couplings $g_1(Q)$ and $g_2(Q)$ is of
particular importance.  In our analysis we use \FS\ 2.1.0
\cite{Athron:2014yba,Athron:2017fvs}, where the $g_i(Q)$ are
determined from the \DR electromagnetic fine-structure constant
$\alpha(Q)$ and the \DR weak mixing angle $\theta_W(Q)$ as
\begin{subequations}
\begin{align}
  g_1(Q) &= \sqrt{\frac{5}{3}}\, \frac{\sqrt{4\pi \alpha(Q)}}{\cos \theta_W(Q)}, \\
  g_2(Q) &= \frac{\sqrt{4\pi \alpha(Q)}}{\sin \theta_W(Q)}.
\end{align}
\end{subequations}
Since version 2.0.0, \FS offers the following two possibilities to
calculate $\theta_W(Q)$:
\begin{enumerate}
\item The $Z^0$ pole mass $\mz$ and the Fermi constant $\gf$ can be
  chosen as input, in which case the weak mixing angle is calculated
  as
  \begin{equation}
    \theta_W(Q) = \frac{1}{2}\arcsin{\sqrt{\frac{2\sqrt{2}\pi \alpha(Q)}{\gf \mz^2 [1 - \Delta\hat{r}(Q)]}}},
    \label{eq:theta_MZGF}
  \end{equation}
  where $\Delta\hat{r}(Q)$ is a function of the one-loop $Z^0$ and the
  $W^\pm$ self-energies \cite{Pierce:1996zz}.

\item The $Z^0$ pole mass $\mz$ and the $W^\pm$ pole mass $\mw$ can be
  chosen as input, in which case the weak mixing angle is calculated
  as
  \begin{equation}
    \theta_W(Q) = \arccos\left(\frac{m_W(Q)}{m_Z(Q)}\right).
    \label{eq:theta_MZMW}
  \end{equation}
  In \eqn{eq:theta_MZMW} $m_W(Q)$ and $m_Z(Q)$ denote the \DR $W^\pm$
  and $Z^0$ masses, respectively, calculated as
  \begin{equation}
    m_V^2(Q^2) = M_V^2 + \frac{1}{(4 \pi)^2}\Re \Pi_V(Q^2, p^2 = M_V^2)
  \end{equation}
  with ${V \in \{ W, Z \}}$ and $\Pi_V$ being the corresponding
  one-loop vector boson self-energy.
\end{enumerate}
Both \eqs{eq:theta_MZGF} and \eqref{eq:theta_MZMW} are equivalent at
the one-loop level, but differ at higher orders.  The difference
involves in particular products of the one-loop vector boson
self-energies.  This difference is of crucial importance for the
existence of multiple solutions at small values of $|\mu|$, as is
shown in \sct{ssec:lightSUSY}.

\subsection{Boundary value problem solvers}

\begin{figure}[t]
  \centering
  \includegraphics[width=0.9\columnwidth]{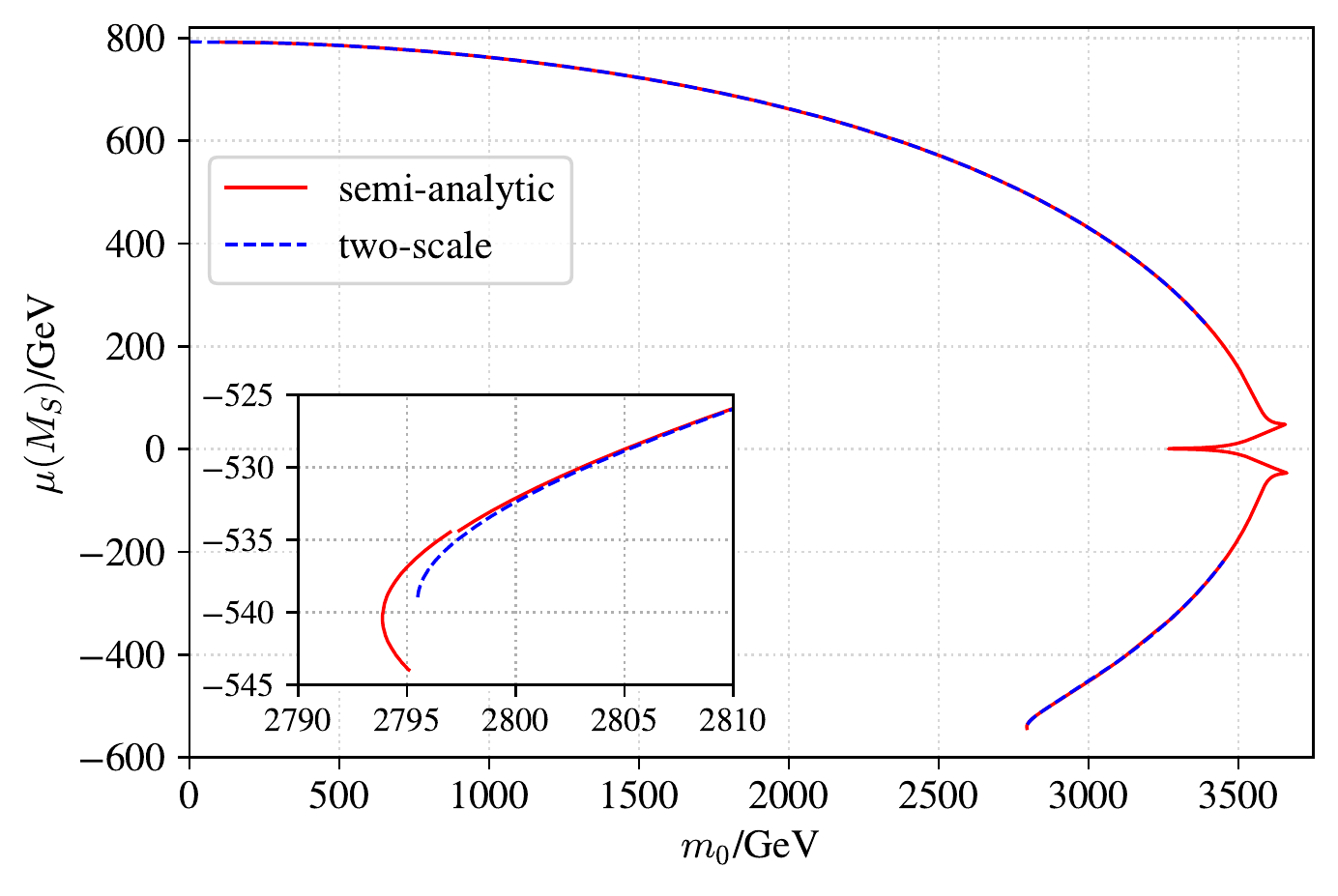}
  \caption{Multiple solutions in the CMSSM, found with the semi-analytic
  solver for the low-energy input parameters $\{ \mz, \gf \}$ and the CMSSM
  parameters given in \eqn{eq:CMSSM_point}. The results from the two-scale
  solver are shown for comparison.}
  \label{fig:CMSSM_TSS_SAS}
\end{figure}

The most common way to solve the CMSSM BVP is the ``two-scale solver''
(TSS), also known as ``running and matching''.  In this approach the
spectrum generator numerically integrates the RGEs and imposes the
boundary conditions at each scale in an iterative way.  If the
iteration converges, the algorithm has found one solution of the
BVP. In the CMSSM one can use the TSS to choose $m_0^2$, $M_{1/2}$,
$A_0$, and $\sign(\mu)$ as input 
and the parameters $\mu$ and $B\mu$ become an output at the SUSY
scale $\ms$.
As an illustration we show in \fig{fig:CMSSM_TSS_SAS} the relation
between $m_0^2$ and $\mu$ for the parameter set
\cite{Allanach:2013cda}
\begin{equation}
  t_\beta(\mz) = 40, \; M_{1/2} = 660 \GeV, \; A_0 = 0,
\label{eq:CMSSM_point}
\end{equation}
and for both signs of $\mu$.  The blue dashed line denotes all points
where the TSS could find a solution.  We see for instance that for
$m_0 = 2\TeV$ the TSS could find a solution for $\sign(\mu) = +1$, but
not for $\sign(\mu) = -1$.  For $m_0 = 3\TeV$ the TSS finds one
solution for each $\sign(\mu)$ and for $m_0 \gtrsim 3.5\TeV$ no
solution is found.

The CNMSSM BVP, however, cannot be solved in this way with the TSS.
There, if one chooses $m_0^2$, $M_{1/2}$, and $A_0$ as input at the
GUT scale, the parameters $\lambda$, $\kappa$ and $v_s$ would need to
be fixed by the EWSB equations at the scale $\ms$.  Since $\lambda$
and $\kappa$ are dimensionless parameters which enter most NMSSM
$\beta$ functions, the iteration between the scales becomes unstable
and the TSS does not converge.

This downside of the TSS led to the development of the semi-analytic
solver (SAS)
\cite{Athron:2009ue,Athron:2009bs,Athron:2011wu,Athron:2012sq,Athron:2012pw,Athron:2017fvs},
which allows one to exchange the role of input and output parameters in constrained models.
To achieve that, the SAS exploits the general structure of the RGEs to
decompose the soft-breaking parameters in terms of GUT parameters
and dimensionless coefficients $c^{(j)}_i$
\cite{Athron:2017fvs}:
\begin{subequations}
\begin{align}
\begin{split}
  m_i^2(Q) &= c^{(1)}_i(Q) m_0^2 + c^{(2)}_i(Q) M_{1/2}^2 \\
  &\phantom{{}={}} \negmedspace + c^{(3)}_i(Q) M_{1/2} A_0 + c^{(4)}_i(Q) A_0^2,
\end{split} \\
  M_i(Q) &= c^{(5)}_i(Q) A_0 + c^{(6)}_i(Q) M_{1/2}, \\
  T_i(Q) &= c^{(7)}_i(Q) A_0 + c^{(8)}_i(Q) M_{1/2}, \\
\begin{split}
  B\mu(Q) &= c^{(9)}(Q) B\mu(\mx) \\
  &\phantom{{}={}} \negmedspace + c^{(10)}(Q) \mu(\mx) M_{1/2} \\
  &\phantom{{}={}} \negmedspace + c^{(11)}(Q) \mu(\mx) A_0.
\end{split}
\end{align}
\label{eq: general SAS}%
\end{subequations}
The coefficients $c^{(j)}_i$ are scale dependent but independent of the
(dimensionful) GUT parameters.  They are determined numerically by
solving the BVP for different values of $m_0^2$, $M_{1/2}$, $A_0$,
and $B\mu(\mx)$ with the TSS. Once the coefficients are known,
\eqs{eq: general SAS} can be solved for the soft-breaking parameters,
which become an output of the algorithm.

The property of the SAS to exchange input and output parameters has
several advantages over the TSS:
\begin{itemize}
\item If the relation between input and output parameters is not
  injective (which occurs in both the CMSSM and CNMSSM), scanning over
  one parameter while obtaining the other one as output, and vice
  versa, allows the search for multiple solutions of the BVP.  This
  procedure is shown by the red solid line in \fig{fig:CMSSM_TSS_SAS},
  where $\mu(\ms)$ is used as input and $m_0^2$ is output.  One
  immediately sees that with the SAS one can find up to four
  solutions for $\mu$ around $m_0 \approx 3.3\TeV$, which have not
  been found by the TSS.  In \sct{sec:CMSSM} we use this
  procedure to study in depth the physical origin of the multiple solutions
  found in \citeres{Allanach:2013yua,Allanach:2013cda,Allanach:2014sea}.
\item In models where the TSS would require dimensionless parameters
  to be output, as for example in the CNMSSM or CE$_6$SSM
  \cite{Athron:2008np,Athron:2009bs}, the SAS enables one to take them
  as input, which yields a stable iteration between the high and low
  scales.  In \sct{sec:CNMSSM} we use this feature in the CNMSSM
  to take the parameters $\lambda(\ms)$, $\kappa(\ms)$, and $v_s(\ms)$
  as input and obtain $m_0^2$, $M_{1/2}$, and $A_0$ as output.
\end{itemize}
On the other hand, there are scenarios where it is advantageous to use
the TSS:
\begin{itemize}
\item If the derivative $\mu'(m_0)$ approaches zero, as happens for example 
  in \fig{fig:CMSSM_TSS_SAS} around $\mu \approx 800 \GeV$, scanning
  over $\mu$ is no longer suitable. Invoking the TSS to vary $m_0$ instead
  and receiving $\mu$ as an output allows to study a region of parameter space
  which cannot be accessed via the SAS as easily.
\item In cases where both solvers find the same unique solution it is a priori 
  not clear whether the TSS or the SAS is the better choice. A combination of 
  both allows a more complete study and a comparison validates the equivalence 
  of the two solvers. In those cases, the TSS in general converges much faster 
  than the SAS does.
\end{itemize}
In the following section we use the semi-analytic approach to
systematically search for multiple solutions to the BVP of the CMSSM
and study their origin in depth.  For our analysis we use the SAS
implemented in \FS\ 2.1.0.
\cite{Athron:2014yba,Athron:2017fvs}.

\section{Multiple solutions in the CMSSM}
\label{sec:CMSSM}

\subsection{Effects from light SUSY particles}
\label{ssec:lightSUSY}
\begin{figure}[t]
  \centering
  \subfigure[]{
  \includegraphics[width = 0.9\columnwidth]{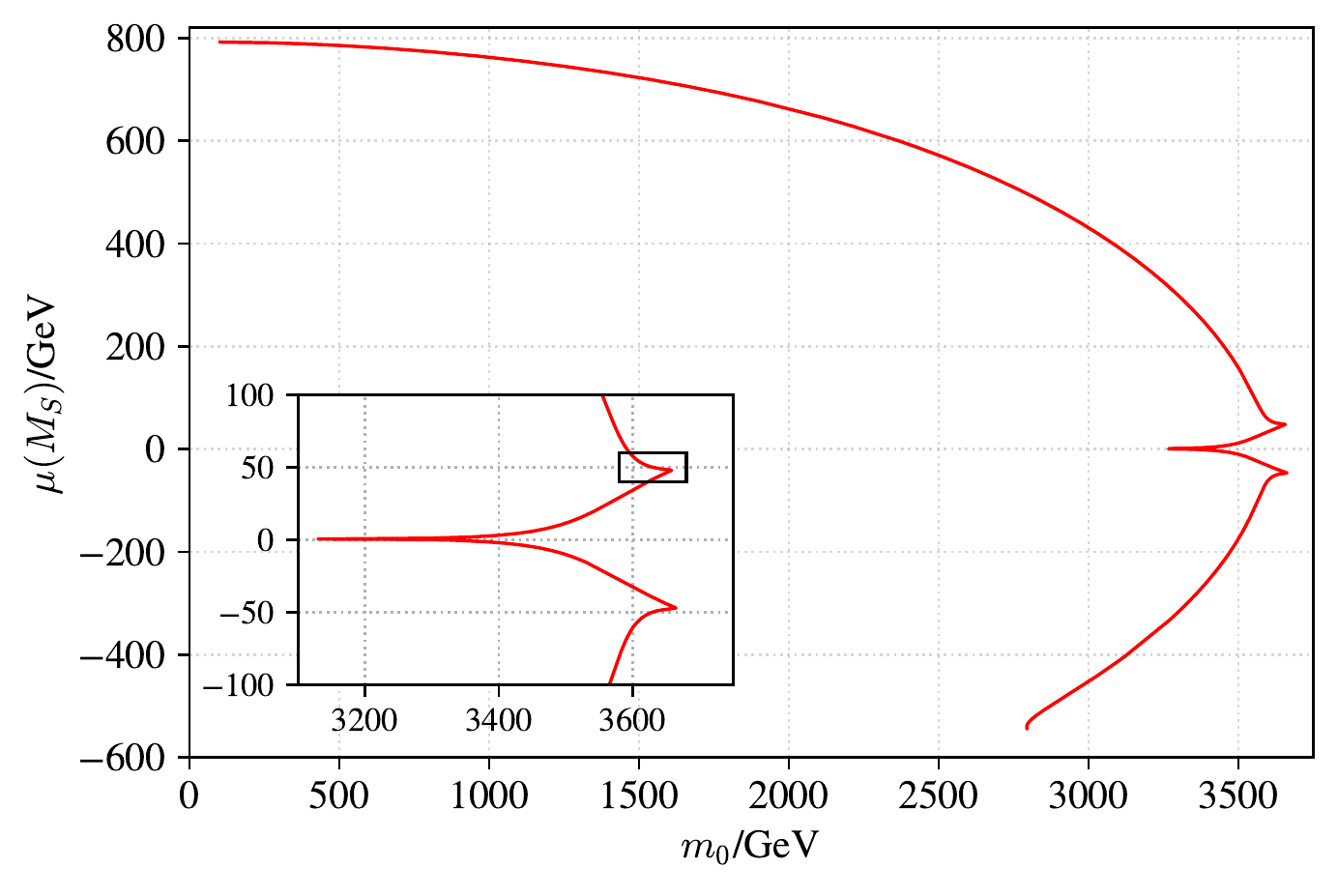}
  \label{fig: mz gf zoom 1}}
  \subfigure[]{
  \includegraphics[width = 0.9\columnwidth]{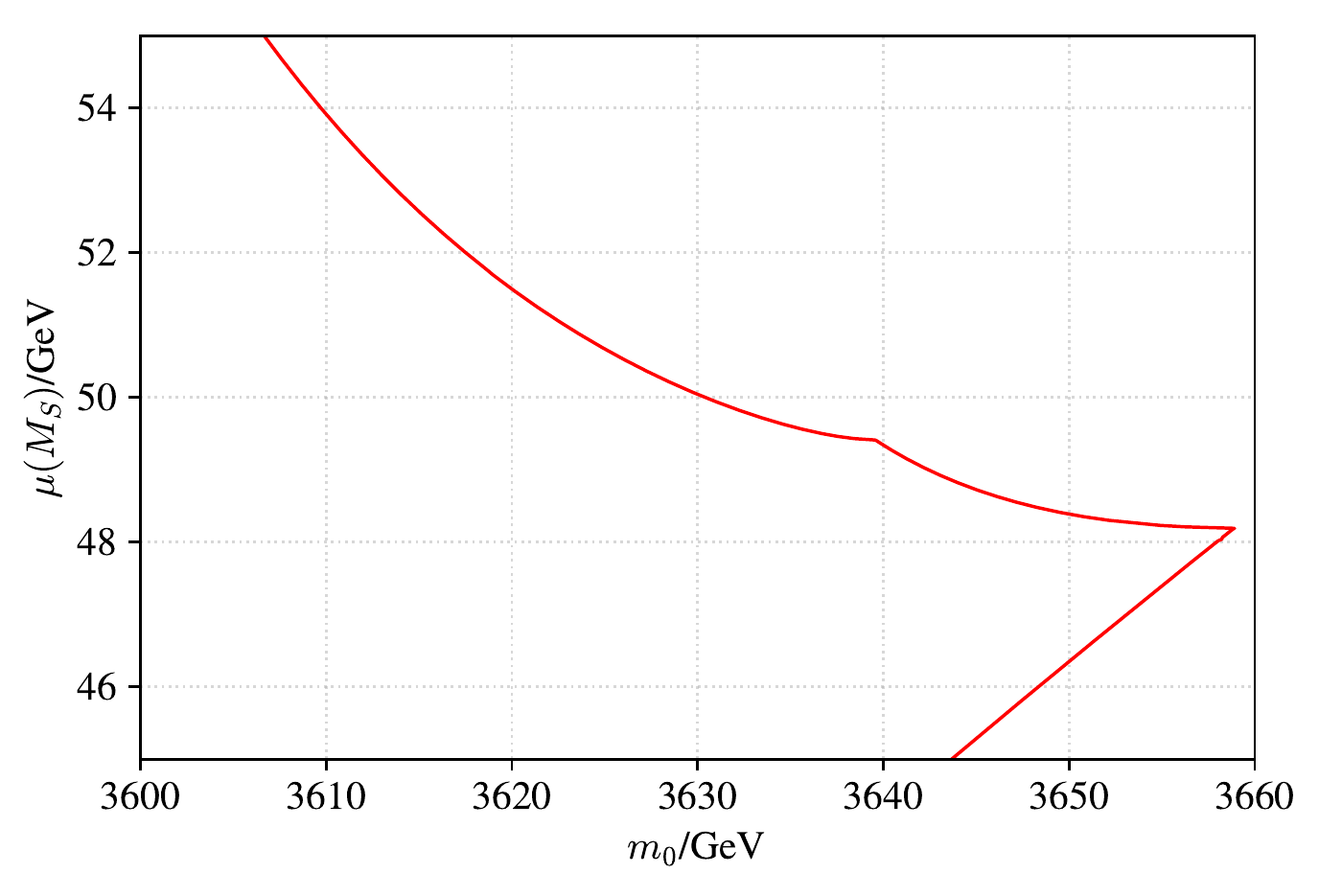}
  \label{fig: mz gf zoom 2}}
  \caption{A scan over $\mu$ with output $m_0$ for the low-energy input
  parameters $\{ \mz, \gf \}$ and the CMSSM parameters given in
  \eqn{eq:CMSSM_point}. In order to make all kinks visible, the boxed
  region in (a) is enlarged and shown in (b).}
  \label{fig: mz gf zoom region I}
\end{figure}

\begin{figure}[t]
  \centering	
  \subfigure[]{
  \includegraphics[width = 0.9\columnwidth]{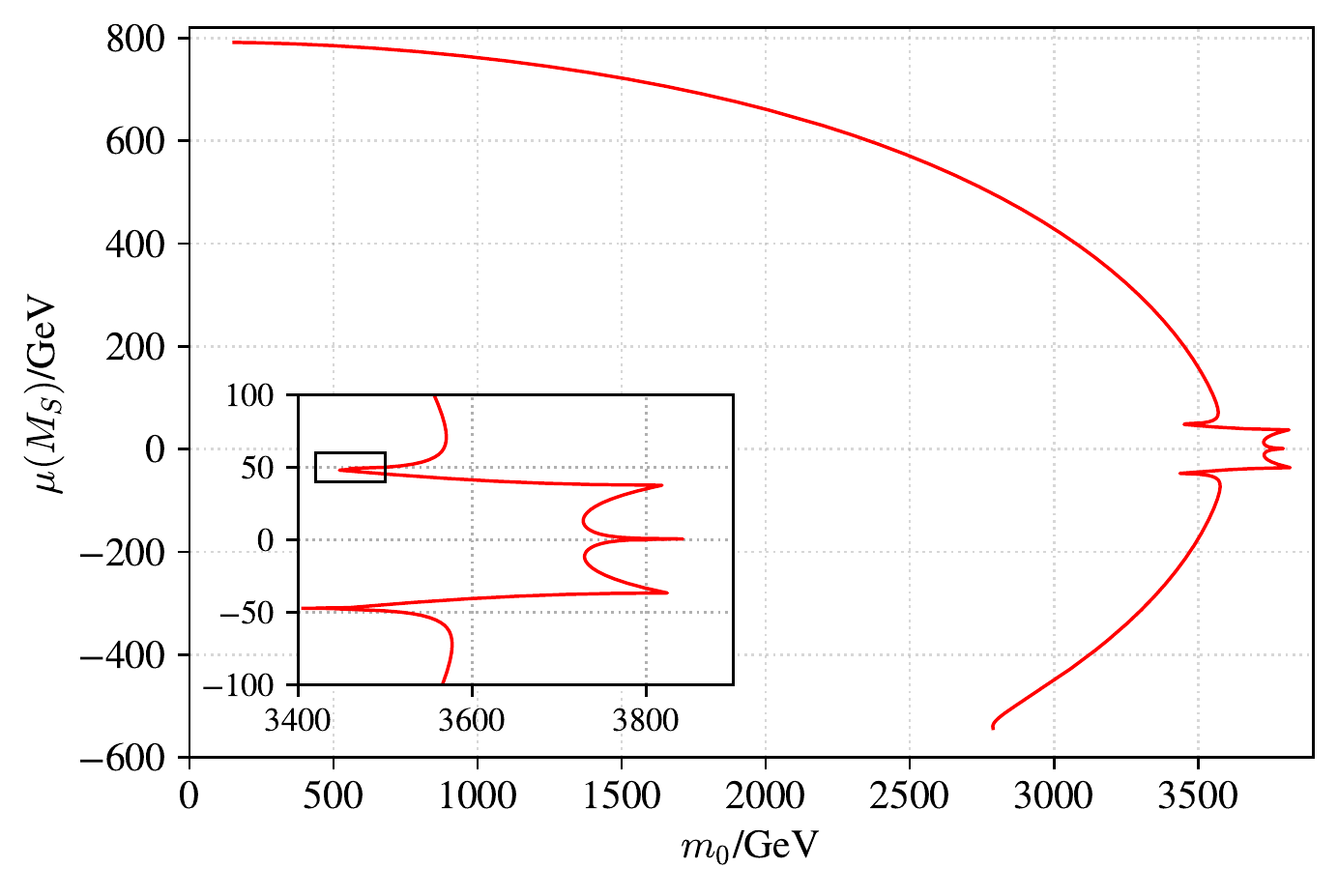}
  \label{fig: mz mw zoom 1}}
  \subfigure[]{
  \includegraphics[width = 0.9\columnwidth]{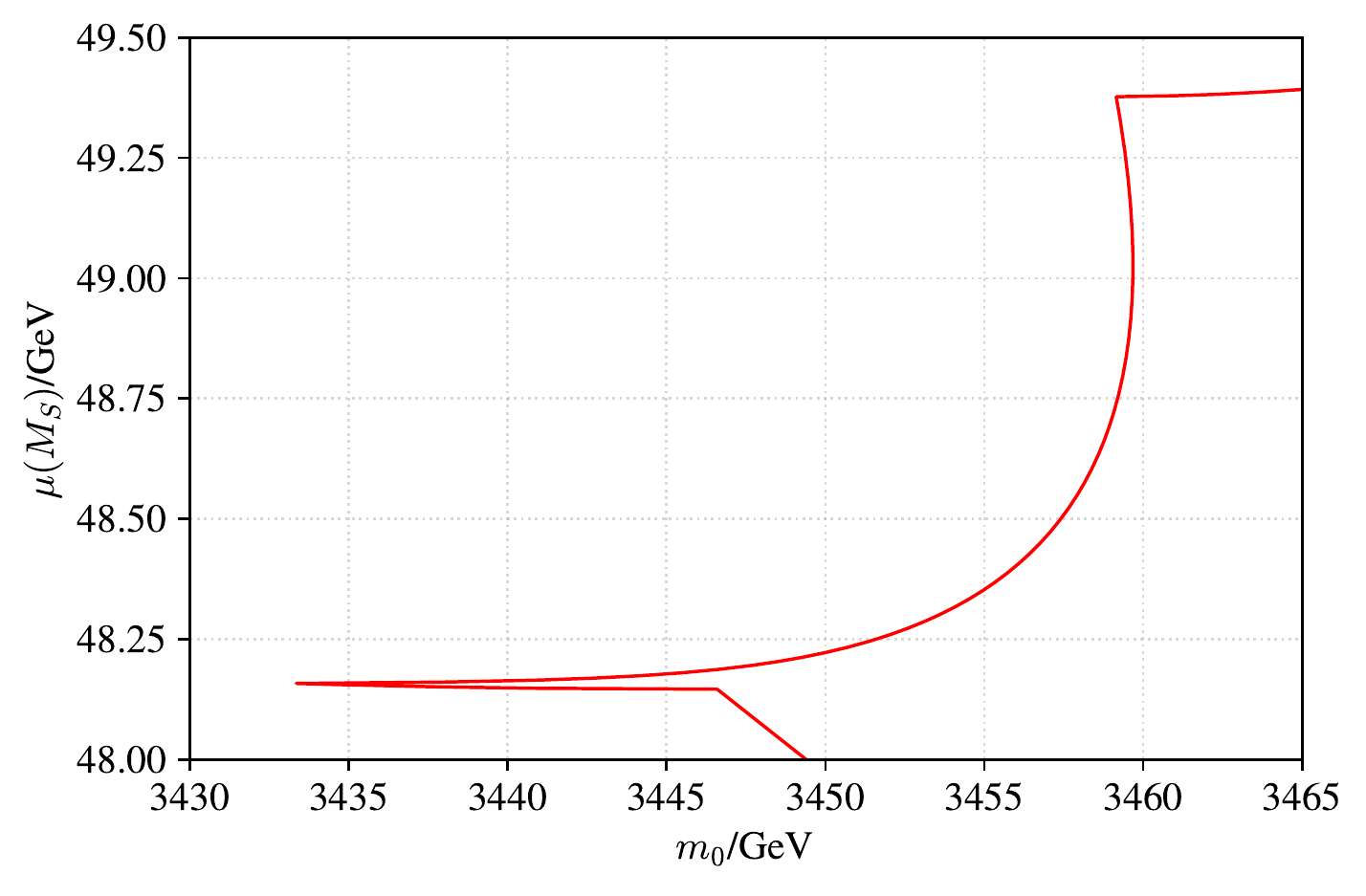}
  \label{fig: mz mw zoom 2}}
  \caption{A scan over $\mu$ with output $m_0$ for the low-energy input
  parameters $\{ \mz, \mw \}$ and the CMSSM parameters given in
  \eqn{eq:CMSSM_point}. In order to make all kinks visible, the boxed
  region in (a) is enlarged and shown in (b).}
  \label{fig: mz mw zoom region I}
\end{figure}

In this section we study the occurrence of multiple solutions in the
CMSSM for small values of $|\mu|$ (see \fig{fig:CMSSM_TSS_SAS}), which
were first observed in \citeres{Allanach:2013cda,Allanach:2013yua}.
Without limitation of generality we restrict our analysis to positive
values of $\mu$ by making use of the approximate mirror symmetry
$m_0(\mu) \approx m_0(-\mu)$, which allows two values for $\mu$ for
fixed $m_0^2$ as long as $|\mu|$ is not too large. As was shown in
\citeres{Allanach:2013cda,Allanach:2013yua}, even for fixed $\sign(\mu)$
the function $m_0(\mu)$ is not necessarily bijective: In the region
$0 < |\mu| \leq 100 \GeV$ the function has several turning points
where the derivative $m_0'(\mu)$ exhibits singular
behavior, see \figs{fig: mz gf zoom region I}, for the
input parameter choice $\{ \mz, \gf \}$. Interestingly, if
$\{ \mz, \mw \}$ is chosen as input, more solutions appear as shown in
\figs{fig: mz mw zoom region I}.  In the following we analyze the
origin of these singularities and derive a mathematical criterion that
describes their positions and count.\footnote{The singularity at
  $\mu = 0$ originates from a massless chargino entering the 1-loop
  threshold correction for $\alpha(M_Z)$.  The region with $\mu = 0$
  is therefore strongly constrained by experimental data and is thus
  not discussed in the following.}

First we consider the case where $\{ \mz, \gf \}$ are chosen as input,
see \figs{fig: mz gf zoom region I}.  In the enlarged subplot of
\fig{fig: mz gf zoom 1} one finds one spike at $\mu\approx 48.2 \GeV$.
Zooming in further reveals one additional kink in \fig{fig: mz gf zoom 2}
at $\mu\approx 49.4 \GeV$.
If $\{ \mz, \mw \}$ are chosen as input (see \figs{fig: mz mw zoom
  region I}), the shape of the curve $m_0^2(\mu)$ is different: In the
zoomed subplot in \fig{fig: mz mw zoom 1} one finds two spikes at
$\mu\approx 37.8 \GeV$ and $\mu\approx 48.16 \GeV$,
respectively. Zooming in further reveals two additional kinks in
\fig{fig: mz mw zoom 2} for $\mu\approx 48.15 \GeV$ and
$\mu\approx 49.4 \GeV$, respectively.
In total one finds two kinks for positive $\mu$ for $\{ \mz, \gf \}$
and four kinks for $\{ \mz, \mw \}$.

\begin{figure}[t]
	\centering
	\includegraphics[width = 0.5\columnwidth]{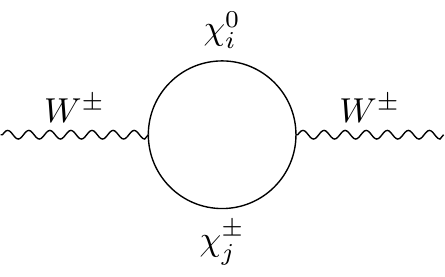}\\
	\includegraphics[width = 0.5\columnwidth]{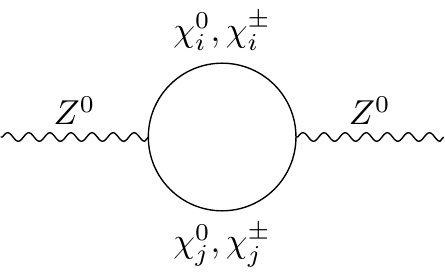}
	\caption{One-loop Feynman diagrams for the vector boson self-energies
	with neutralinos and charginos in the loop.}
	\label{fig: Feynman diagrams}
\end{figure}

The origin of these kinks can be traced back to singular chargino and
neutralino contributions to the one-loop vector boson self-energies
$\Pi_Z$ and $\Pi_W$ (see \fig{fig: Feynman diagrams}), which are used
to calculate the electroweak gauge couplings at the scale $Q = \mz$,
as described in \sct{sec:CMSSM_BPV}.  Since the gauge couplings contribute
to every $\beta$ function of the MSSM, a singular point in
$\Pi_V'(\mu)$ translates into a singular point in $m_0'(\mu)$.  Note
that the precise dependence of the $g_i(Q)$ on the $\Pi_V$ depends on
the chosen set of input parameters, $\{ \mz, \gf\}$ or $\{ \mz, \mw \}$.
This explains the different singularities between \figs{fig: mz gf
  zoom region I} and \ref{fig: mz mw zoom region I}.
In the following we describe the intricate $\mu$ dependence of $m_0$
and investigate the source of the singularities of $m_0'(\mu)$.

The vector boson self-energies $\Pi_V$ depend on $\mu$ via the
chargino and neutralino masses, which enter the one-loop
diagrams shown in \fig{fig: Feynman diagrams}.  Expressed in terms the
loop functions $H_0$ and $B_0$ \cite{Pierce:1996zz}, each diagram
gives a contribution
\begin{equation}
\begin{aligned}
  \Pi_V(Q^2, p^2) &\supset 4 m_1m_2 \Re \big( C_L^* \, C_R \big) B_0(p^2, m_1^2, m_2^2) \\
  &\phantom{{}\supset{}} \negmedspace +\big( \left| C_L \right|^2 + \left| C_R \right|^2 \big)
  H_0(p^2, m_1^2, m_2^2) 
\end{aligned}
\label{eq: self-energy general}
\end{equation}
to the self-energies. In \eqn{eq: self-energy general} $p^2$ denotes
the external momentum squared, $m_{1, 2}$ are the running masses of
the fermions in the loop, and $C_L$ and $C_R$ are vertex coefficients
derived from the Lagrangian. All running quantities are evaluated at
the renormalization scale $Q = \mz$. In total there are fourteen
electroweakino diagrams contributing to $\Pi_Z$ and eight to
$\Pi_W$. Of these, four ($Z_0$) and two ($W^\pm$) contain only
higgsino-like charginos and neutralinos, whose masses are
approximately given by $|\mu|$.  Therefore, only those six diagrams
are responsible for the kinks.

The singularities in $\Pi_V'(\mu)$ now have the following deeper origin: In
the \DR renormalization scheme the finite part of the $B_0$ function
appearing in \eqn{eq: self-energy general} is given by
\begin{equation}
  B_0(p^2, m_1^2, m_2^2) = -\ln\left( \frac{p^2}{Q^2} \right) - f_B(x_+) - f_B(x_-),
  \label{eqn: B0 ev}
\end{equation}
where
\begin{subequations}
\begin{align}
  &f_B(x) = \ln(1-x) - x\ln(1-x^{-1}) - 1, \\
  &x_\pm = \frac{s \pm \sqrt{s^2 - 4p^2(m_1^2-\text{i}\epsilon)}}{2p^2} \label{eq: radicand}, \\
  &s = p^2 - m_2^2 + m_1^2.
\end{align}
\end{subequations}
\begin{figure}[t]
  \centering
  \includegraphics[width = 0.9\columnwidth]{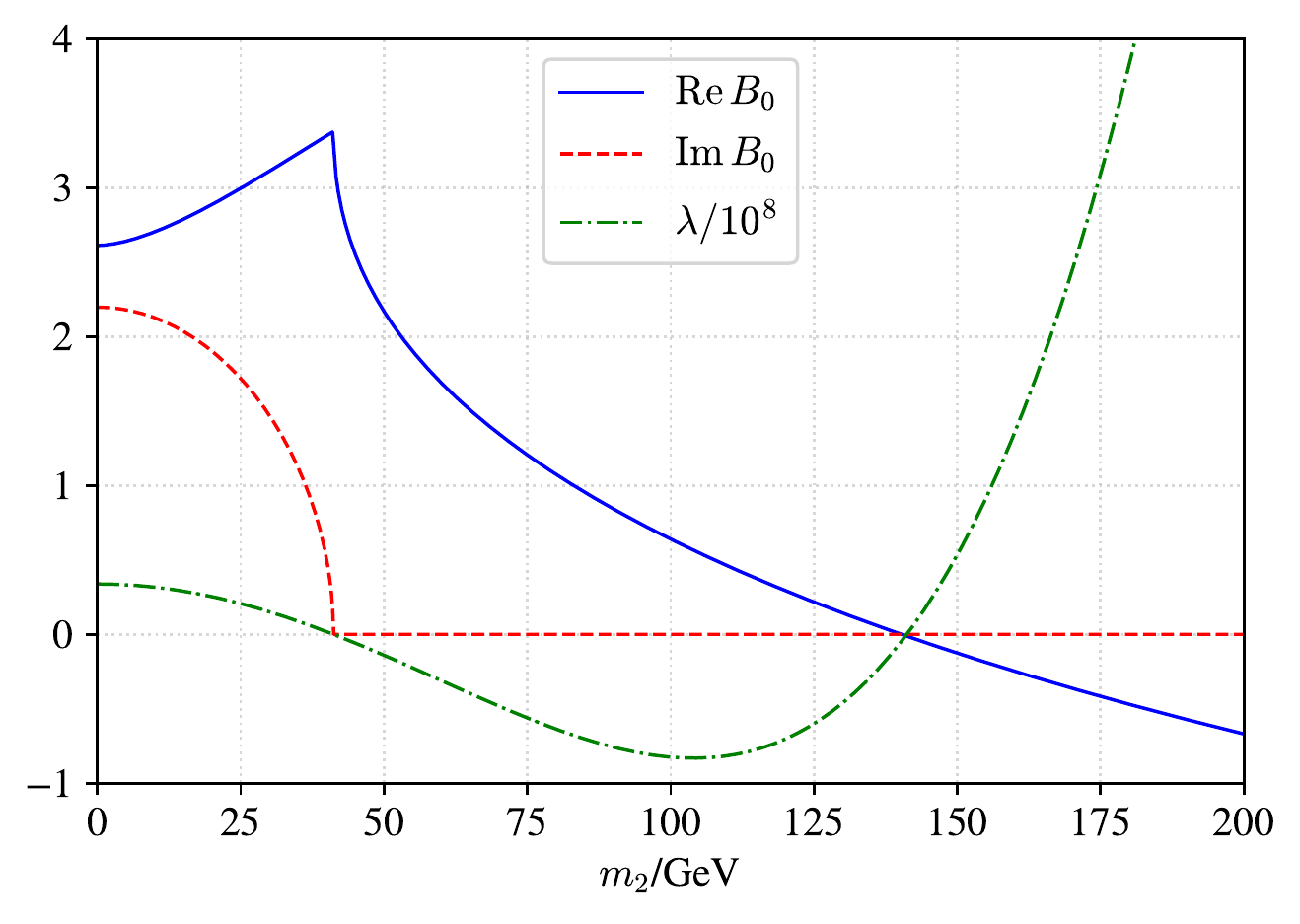}
  \caption{Real and imaginary part of $B_0(p^2,m_1^2,m_2^2)$ for
  masses $m_1 = 50 \GeV$ and $m_2 \in [0, 200] \GeV$. The external
  momentum $p$ and the renormalization scale $Q$ are both set to $\mz$.
  The \kallen function $\lambda(p^2,m_1^2,m_2^2)$ is shown for comparison.}
  \label{fig: B0 r and i}
\end{figure}%
In \fig{fig: B0 r and i} the real and imaginary part of $B_0$ are shown
exemplary as a function of the mass $m_2$ for fixed $m_1$.  One finds
that as soon as $m_2$ becomes small enough such that both particles in
the loop go on-shell simultaneously, $B_0$ acquires an imaginary part
and the derivatives of the real and the imaginary part are singular for
this value of $m_2$. This is due to the square root function in
\eqn{eq: radicand} not being differentiable when its argument
vanishes. The radicand, which is just the \kallen function%
\footnote{$\lambda(a,b,c) = a^2+b^2+c^2-2ab-2ac-2bc$}
$\lambda(p^2, m_1^2, m_2^2)$, has two roots
\begin{equation}
  p^2 - m_1^2 - m_2^2 = \pm 2m_1m_2,
\label{eq: Kallen zero}
\end{equation}
of which the one with positive sign is located at $m_2 \approx 41 \GeV$
and the one with negative sign is located at $m_2 \approx 141 \GeV$ in
\fig{fig: B0 r and i}. Both real and imaginary part are not differentiable at 
the zero $m_2 \approx 41 \GeV$, but they are at the other due to a 
cancellation between the $f_B$-terms in \eqn{eqn: B0 ev}. The appearance
of an imaginary part is in accordance with the optical theorem from which
we expect the self-energy to become complex when the particles in the
loop are light enough to go on-shell simultaneously.

With this knowledge one would expect four higgsino-like (and thus
$\mu$-dependent) combinations ($\chi_1^0 \chi_1^0$,
$\chi_1^0 \chi_2^0$, $\chi_2^0 \chi_2^0$, $\chi_1^\pm \chi_1^\pm$) to
give rise to singularities in the $Z^0$ self-energy, and two
combinations ($\chi_1^0 \chi_1^\pm, \chi_2^0 \chi_1^\pm$) for the
$W^\pm$ self-energy. Following, there should be four spikes for the
input parameters $\{ \mz, \gf \}$ and six for $\{ \mz, \mw \}$ in
total. However, there are only two and four spikes in $m_0(\mu)$,
respectively.  Evaluating \eqn{eq: self-energy general} in the limit
of a vanishing \kallen function renders it in a form in which the
disappearance of some singularities becomes manifest. To this end we
use the identities
\begin{subequations}
\begin{align}
  &H_0 = 4 B_{22} + G_0, \\
  &B_{22} \supset -\frac{1}{12p^2} \lambda (p^2, m_1^2, m_2^2) B_0, \\ 
  &G_0 \supset ( p^2 - m_1^2 - m_2^2 ) B_0,
\end{align}
\label{eq: H0s}%
\end{subequations}
where we have neglected constants as well as terms which are
proportional to the one-point function $A_0$ \cite{Pierce:1996zz}.
Plugging \eqs{eq: H0s} in \eqn{eq: self-energy general} and setting
$\lambda = 0$, we arrive at
\begin{equation}
\begin{aligned}
  \Pi_V \big|_{\lambda = 0} &\supset \big[ ( p^2 - m_1^2 - m_2^2 )
  \big( \left| C_L \right|^2 + \left| C_R \right|^2 \big) \\
  &\phantom{{}\supset{}} \negmedspace + 4m_1m_2 \Re \big( C_L^* C_R \big) \big]
  B_0(p^2, m_1^2, m_2^2).
\end{aligned}
\label{eq: self-energy 2}
\end{equation}
We plug the root \eqn{eq: Kallen zero} with positive sign into \eqn{eq: self-energy 2} to obtain
\begin{equation}
\begin{aligned}
  \Pi_V \big|_{\lambda=0} &\supset \big[ 2m_1m_2 \big( \left| C_L \right|^2 + 
  \left| C_R \right|^2 \big) \\
  & \phantom{{}\supset{}} \negmedspace + 4m_1m_2 \Re \big( C_L^* C_R \big) \big]
  B_0(p^2, m_1^2, m_2^2) \\
  &= 2m_1m_2 \left| C_L + C_R \right|^2 B_0(p^2, m_1^2, m_2^2).
\end{aligned}
\label{eq: self-energy Kallenlimit}
\end{equation}
\eqn{eq: self-energy Kallenlimit} is the non-vanishing part of a self-energy graph 
in the limit where the \kallen function $\lambda(p^2, m_1^2, m_2^2)$ is zero. 
From this we infer that whenever ${C_L + C_R}$ is zero in the limit ${\lambda \to 0}$, 
the corresponding diagram does not contribute a singularity to $\Pi_V$.

Since the entries of the chargino mixing matrices $U^-$ and $U^+$ are
independent, there is no relation which would guarantee the
cancellation $C_L + C_R = 0$, as long as at least one chargino appears
in the loop. Hence, the diagrams with $\chi_1^0 \chi_1^\pm$,
$\chi_2^0 \chi_1^\pm$, and $\chi_1^\pm \chi_1^\pm$ in the loop
contribute non-vanishing singularities to $\Pi_V$.

The other three relevant diagrams are pure $\chi_i^0\chi_j^0$
contributions to the $Z^0$ self-energy. For a
$Z^0\chi_i^0\chi_j^0$ vertex the relation $C_L^* = -C_R$ holds
\cite{Pierce:1996zz}. Since the $C_{L/R}$ are in general complex
quantities, this property is not sufficient for $C_L + C_R$ to
vanish. If the neutralinos are identical, however, $C_L$ and $C_R$
become real and cancel when added.

We conclude that the $Z^0$ self-energy diagrams with identical
neutralinos in the loop do never give a spike, since the singular terms
in $\Pi_Z$ have a vanishing coefficient in the limit $\lambda \to
0$. The singularities (kinks or spikes) originate only from diagrams with light
$\chi_1^0 \chi_1^\pm$, $\chi_2^0 \chi_1^\pm$, $\chi_1^\pm \chi_1^\pm$,
or $\chi_1^0 \chi_2^0$ in the loop. 

Note, that at the singular points spikes or kinks can appear in the
overall vector boson self-energies.  However, only spikes lead to
multiple solutions, because there the sign of the derivative around
the singular point changes, which results in a turning point.
Whether a singularity causes a spike or a
kink depends on the relative sign between the $B_0$ function which
causes the singularity and other loop corrections to the self-energy.
The relative signs generally depend on the regarded model as well as
on the input parameters.

\subsection{Effects from non-linear parameter inter-dependencies}
\label{ssec:nonlinear_effects}
\begin{figure}[t]
  \centering
  \includegraphics[width = 0.9\columnwidth]{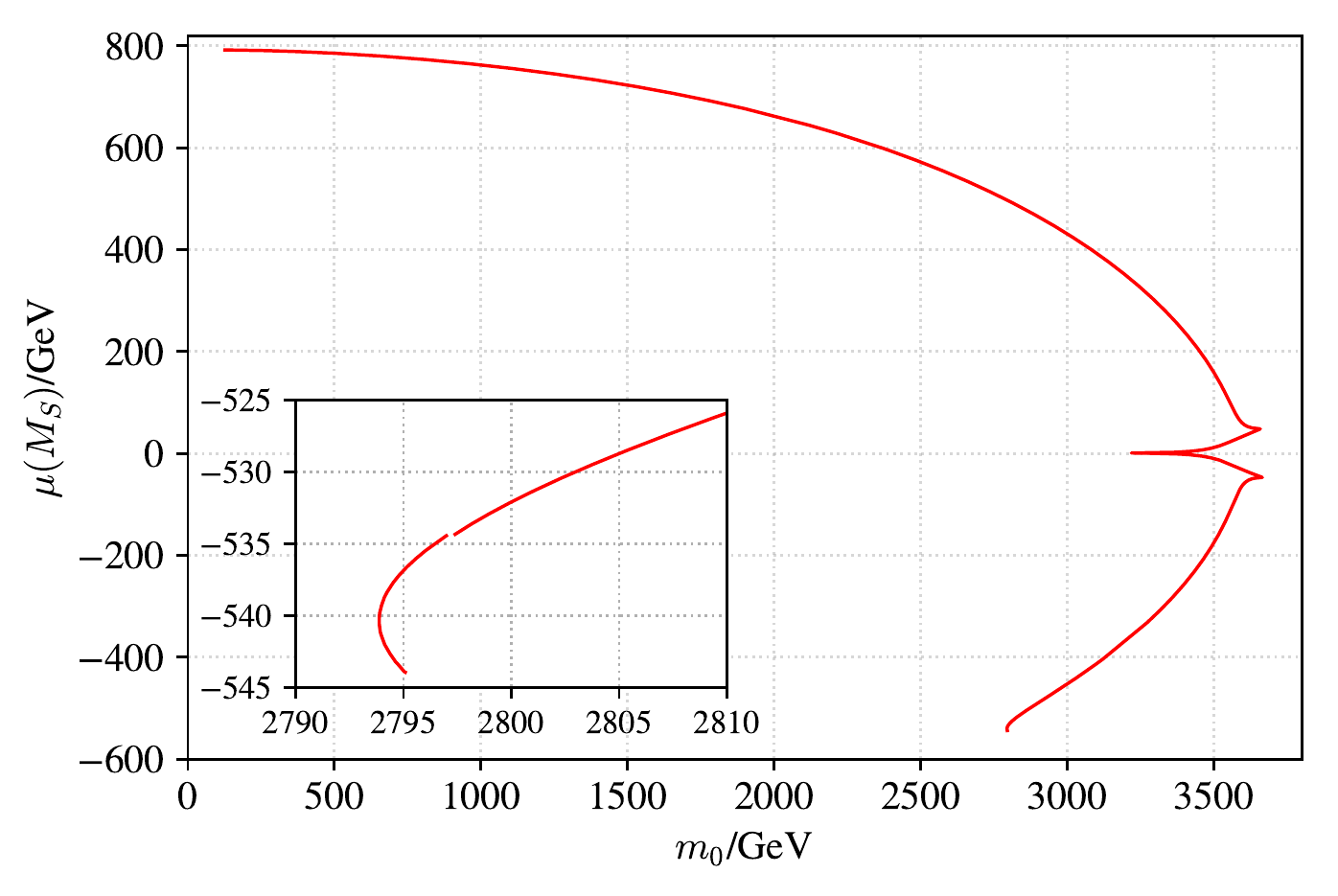}
  \caption{The SAS results with zoom for the low-scale input parameters
  $\{ \mz, \gf \}$ and the CMSSM parameters from \eqn{eq:CMSSM_point}.
  $m_0(\mu)$ has a minimum at $\mu = -540.5 \GeV$, which gives rise
  to multiple solutions to the CMSSM BVP.}
  \label{fig: zoom region II}
\end{figure}

In \citeres{Allanach:2013cda,Allanach:2013yua} another CMSSM
parameter region with multiple solutions was found around
$\mu \approx -500 \GeV$, see \fig{fig: zoom region II}.  In this
section we investigate this parameter region with multiple solutions,
which we find are not caused by light particles in vector boson
self-energies, but by intricate nonlinear parameter inter-dependencies.

The non-monotonic behavior of $m_0(\mu)$ in the region
$\mu \approx -500 \GeV$ is depicted in \fig{fig: zoom region II}.
Multiple solutions appear in this region, because the function
$m_0'(\mu)$ changes its sign such that $m_0(\mu)$ has a minimum at
$\mu = -540.5 \GeV$.  For too small values of
$\mu \lesssim -545 \GeV$ there is no physical solution because the running
masses of the neutral Higgs bosons become tachyonic in this region of 
parameter space. In the following we study the parameter
interplay which is responsible for the existence of the minimum.

To this end, we derive an approximate relation between $m_0$ and $\mu$.
Our starting point is the tree-level EWSB equation
\begin{equation}
  \mu^2 = \left(m_{h_d}^2 t_\beta^{-1}-m_{h_u}^2 t_\beta \right)
  \frac{t_\beta}{t_\beta^2-1}-\frac{1}{2} m_Z^2,
\end{equation}
which is imposed at the SUSY scale $\ms$. Hence, all
renormalization scale dependent quantities are evaluated at $\ms$.
For our set of input parameters, $t_\beta^2 \gg 1$ holds and we can
also neglect the $m_Z$-term, which implies
\begin{equation}
  \mu^2 = m_{h_d}^2 t_\beta^{-2}-m_{h_u}^2.
\label{eqn: EWSBapprox}
\end{equation}
To relate those parameters to $m_0^2$, we make use of the RGEs
\begin{equation}
  m_{h_i}^2 = m_0^2 + \beta_{m_{h_i}^2} \log \frac{\ms}{\mx}, \, i \in \{ u,d \}.
\label{eqn: mHi running}
\end{equation}
At leading order the $\beta$ functions are independent of $\mu$ and  fulfill the
hierarchy $\beta_{m_{h_u}^2} > \beta_{m_{h_d}^2}$. This hierarchy arises 
because the up-type $\beta$ function is proportional to the squared top-Yukawa coupling whereas 
the down-type one contains only down-type fermion contributions. It causes $m_{h_d}^2$ to be
close to $m_0^2$ and at the same time allows $m_{h_u}^2$ to become negative---which
is usually necessary for EWSB to occur. We will therefore only replace $m_{h_d}^2$ in 
\eqn{eqn: EWSBapprox} by \eqn{eqn: mHi running} and arrive at
\begin{equation}
  m_0^2 = t_\beta^2 \left( \mu^2+m_{h_u}^2 \right) + \beta_{m_{h_d}^2} \log \frac{\mx}{\ms}.
\end{equation}
This, to leading order, implies that a minimum of $m_0^2$ appears when
\begin{equation}
  \frac{\text{d} m_{h_u}^2}{\text{d} \mu} = -2\mu.
\label{eqn: mHu2_mu}
\end{equation}
For our choice of parameters, a numerical analysis leads to
\begin{equation}
  \frac{\text{d} m_{h_u}^2}{\text{d} \mu} \bigg|_{\mu = -540.5 \GeV}
  \approx 1081 \GeV,
\end{equation}
so \eqn{eqn: mHu2_mu} is fulfilled for $\mu=-540.5 \GeV$. A comparison
with \fig{fig: zoom region II} validates that this indeed is the position of
the minimum around which multiple solutions occur. For differently chosen
input parameters the existence of a point where the derivative of
$m_{h_u}^2$ is of the appropriate size to create a minimum is not
ensured. An example for this is given in \sct{ssec:CMSSM_excluded}.

\subsection{Multiple solutions for $M_{1/2}$ and $A_0$}
\label{ssec:different_output_pars}

\begin{table}[b]
  \caption{CMSSM input/output parameters}
  \label{tab:CMSSM_parameters}
  \begin{tabular}{lll}
    \hline\noalign{\smallskip}
    solver & input & output  \\
    \noalign{\smallskip}\hline\noalign{\smallskip}
    TSS & $m_0^2$, $M_{1/2}$, $A_0$, $t_\beta$ & $\mu$, $B\mu$ \\
    SAS1 & $\mu$, $M_{1/2}$, $A_0$, $t_\beta$ & $m_0^2$, $B\mu$ \\
    SAS2 & $\mu$, $m_0^2$, $A_0$, $t_\beta$ & $M_{1/2}$, $B\mu$ \\
    SAS3 & $\mu$, $m_0^2$, $M_{1/2}$, $t_\beta$ & $A_0$, $B\mu$ \\
    \noalign{\smallskip}\hline
  \end{tabular}
\end{table}

The SAS allows one to exchange input and output parameters of
constrained (SUSY) models.  This property was used in the previous
sections to exchange the role of $m_0^2$ and $\mu$ in the CMSSM to
examine the function $m_0^2(\mu)$ for (non-)injectivity.  However, the
SAS is not restricted to the exchange $m_0^2 \leftrightarrow \mu$
(SAS1) and further choices are possible, see
\tab{tab:CMSSM_parameters}.  In the following we apply the SAS to
study the relations $M_{1/2} \leftrightarrow \mu$ (SAS2) and
$A_0 \leftrightarrow \mu$ (SAS3) to investigate the implications of
their non-bijectivity.

\begin{figure}[t]
\centering
\subfigure[$m_0 = 3000 \GeV$]{
\includegraphics[width = 0.85\columnwidth]{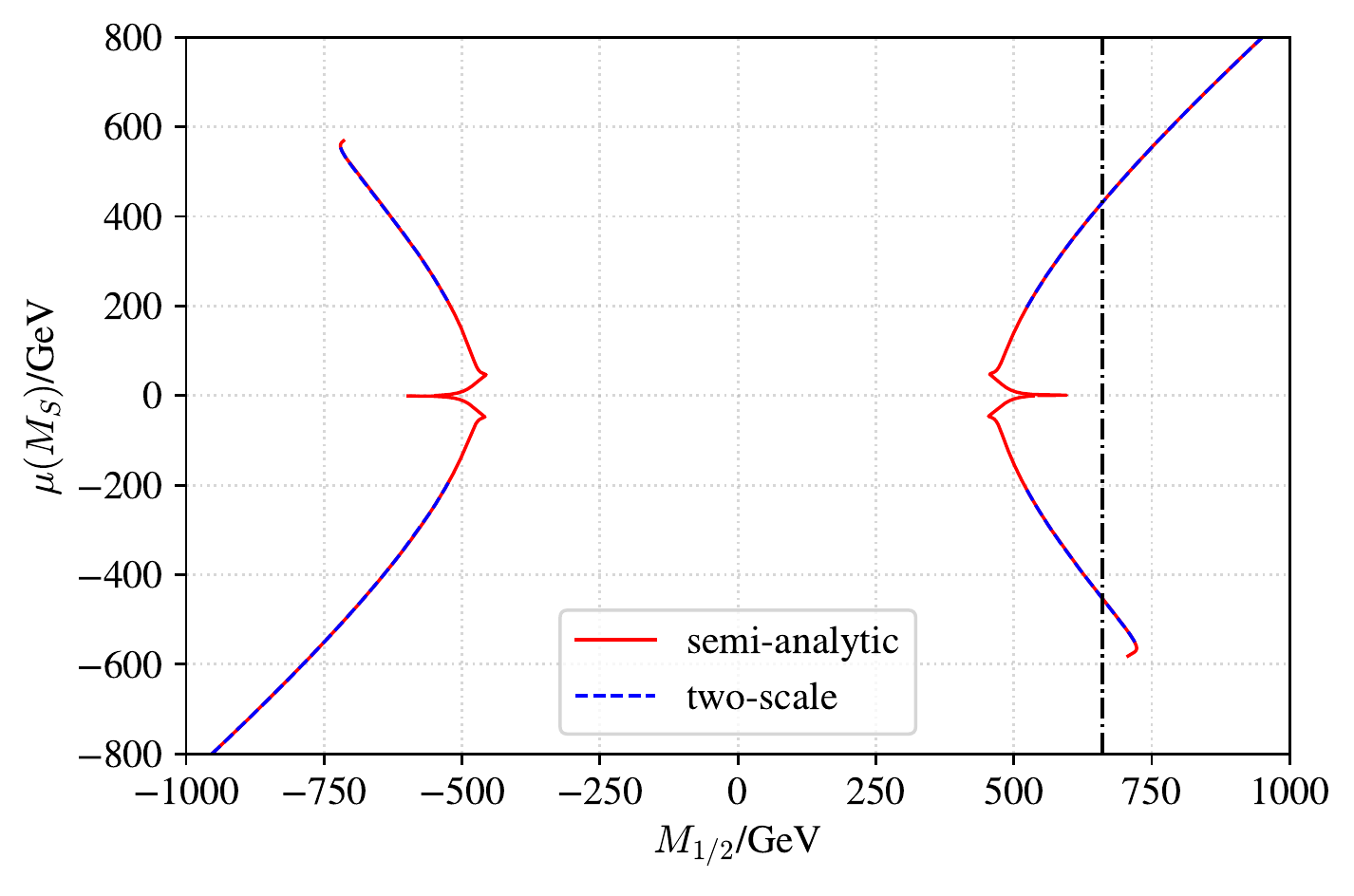}
\label{fig: M12 out 3000}}\\
\subfigure[$m_0 = 3500 \GeV$]{
\includegraphics[width = 0.85\columnwidth]{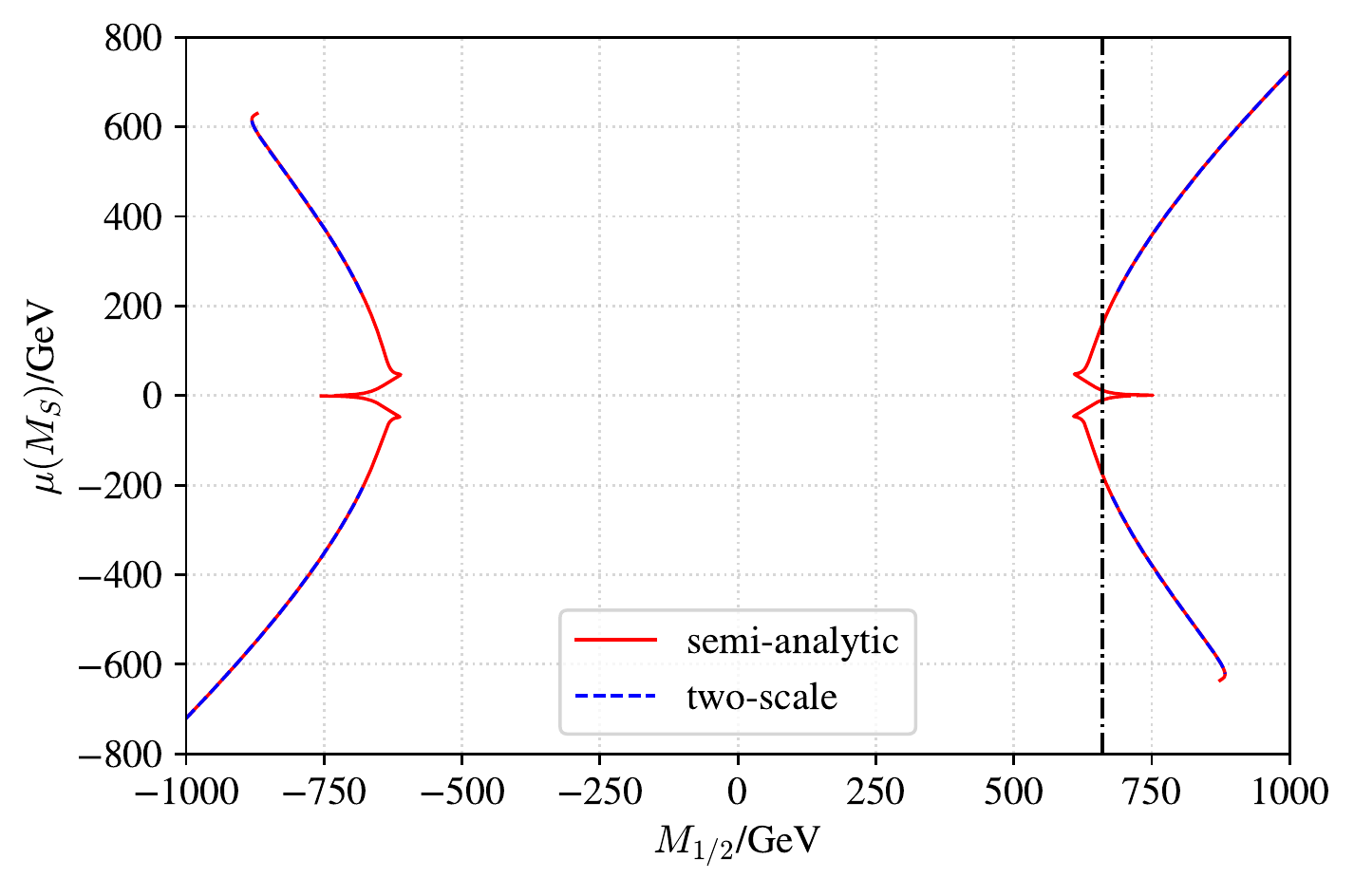}
\label{fig: M12 out 3500}}
\caption{The relation between $M_{1/2}$ and $\mu$ obtained with the TSS 
  (blue) and the SAS2 (red) for two different values of $m_0$.  The dash-%
  dotted lines (black) give the points where $M_{1/2} = 660 \GeV$.  The
  intersection points of the colored and the black lines mark the same
  solutions as the ones shown in \fig{fig:CMSSM_TSS_SAS}.}
\label{fig:_M12_out}
\end{figure}
The semi-analytic solver can treat the GUT parameter $M_{1/2}$ as
output and $\mu$ as input (SAS2) by solving the EWSB equations
for $M_{1/2}$.  To do so, one makes use of the semi-analytic ansatz
\eqs{eq: general SAS} again.  The equation involving the $m_{h_i}$
is quadratic in $M_{1/2}$, in contrast to being linear in $m_0^2$,
which allows for up to two distinct solutions for fixed $\mu$.
In \figs{fig: M12 out 3000} and \ref{fig: M12 out 3500} we scan over
$\mu$ (SAS2) and $M_{1/2}$ (TSS) for $A_0 = 0$, $t_\beta = 40$
and two different values $m_0 \in \{ 3000, 3500 \} \GeV$.
As expected, the semi-analytic solver finds two branches, one for
each solution of the quadratic equation. The TSS only finds solutions 
which are obtained from SAS2 as well. The solutions at
$M_{1/2} = 660 \GeV$ (vertical black lines) are the same as found in
\sct{ssec:lightSUSY} using SAS1. Thus, scanning over $\mu$ while
using $M_{1/2}$ as an output parameter does not give new solutions
compared to the case where $m_0^2$ is output.

\begin{figure}[t]
\centering
\subfigure[$m_0 = 3000 \GeV$]{
\includegraphics[width = 0.85\columnwidth]{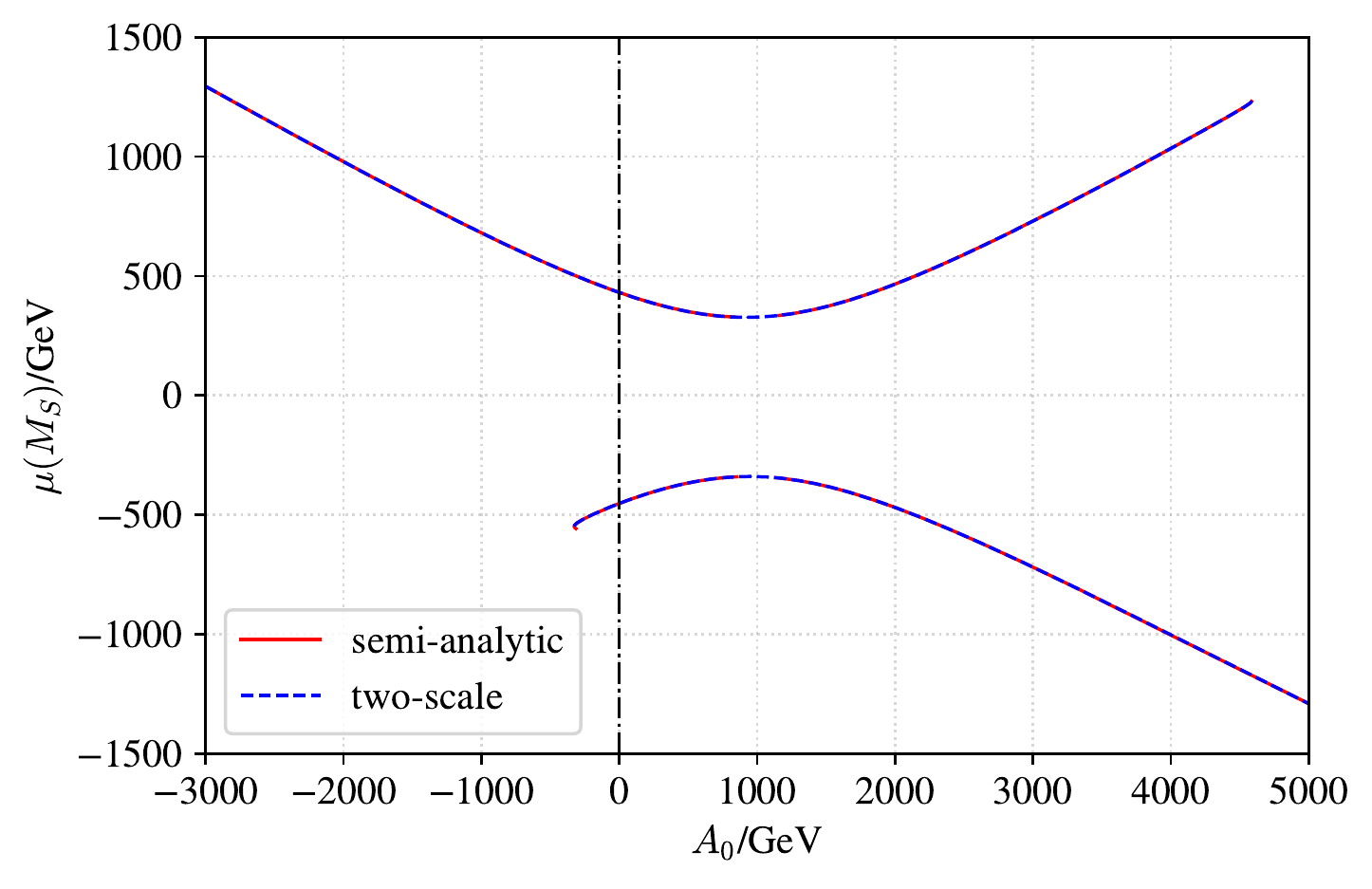}
\label{fig: A0 out 3000}}\\
\subfigure[$m_0 = 3500 \GeV$]{
\includegraphics[width = 0.85\columnwidth]{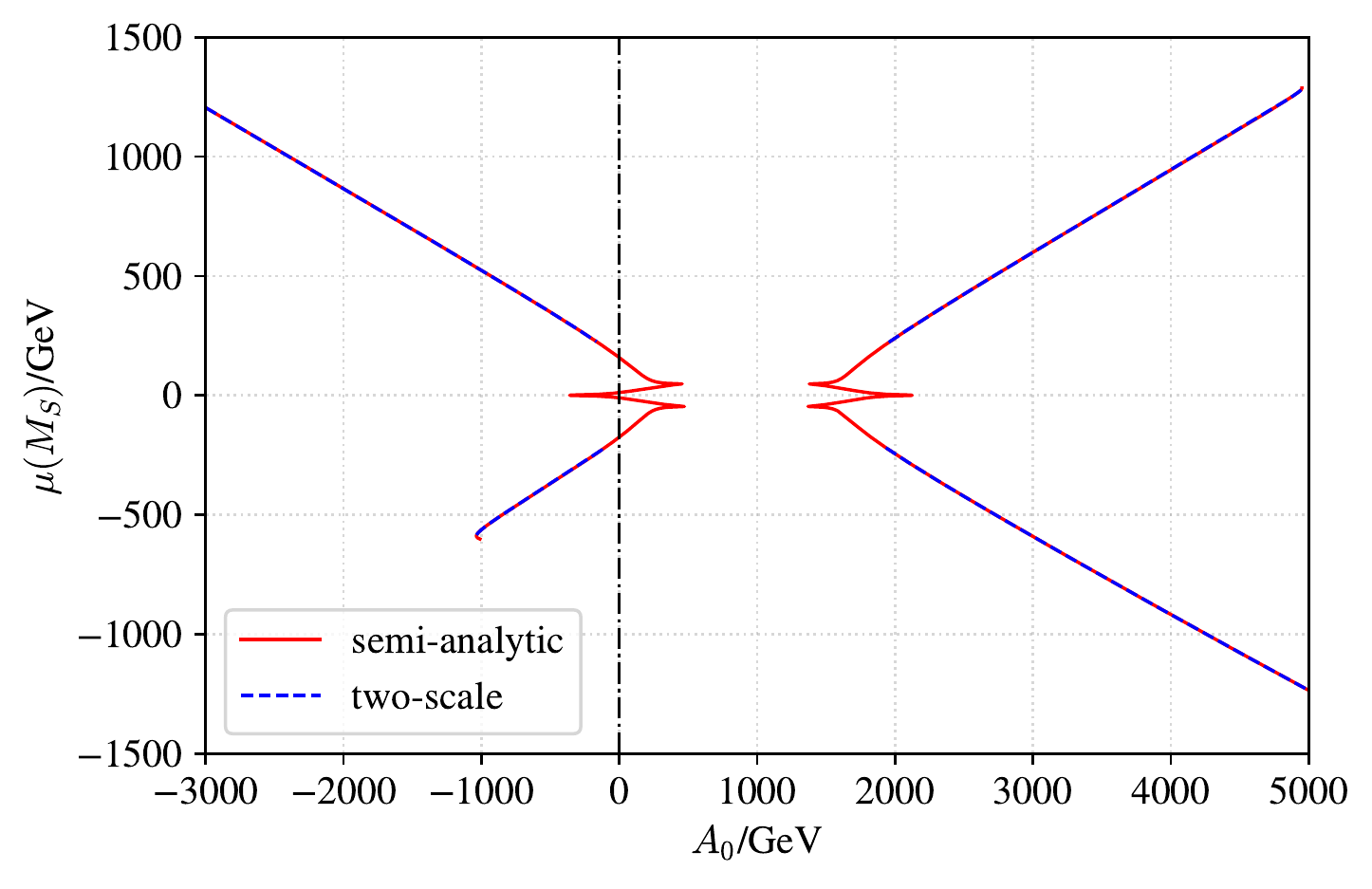}
\label{fig: A0 out 3500}}
\caption{The relation between $A_0$ and $\mu$ obtained with the TSS 
  (blue) and the SAS3 (red) for two different values of $m_0$.  The dash-%
  dotted lines (black) give the points where $A_0 = 0$.  The intersection
  points of the colored and the vertical black lines mark the same
  solutions as the ones shown in \fig{fig:CMSSM_TSS_SAS}.}
\label{fig:_A0_out}
\end{figure}
Similarly to SAS2, $A_0$ can be treated as output parameter by solving
the EWSB equations for $A_0$ (SAS3) instead.  Also in this case the
ansatz is quadratic in $A_0$, so up to two distinct solutions are
possible.
In \figs{fig: A0 out 3000} and \ref{fig: A0 out 3500} we show $A_0(\mu)$
and $\mu(A_0)$ for $M_{1/2} = 660 \GeV$, $t_\beta = 40$, and
$m_0 \in \{ 3000, 3500 \} \GeV$.  We find that for $m_0 = 3500 \GeV$
(\fig{fig: A0 out 3500}) the SAS can scan over both solution branches,
while for $m_0 = 3000 \GeV$ (\fig{fig: A0 out 3000}) the two branches
have merged. The multiple solutions around $\mu \approx 0$ have
vanished and a region without solutions ($|\mu| \lesssim 300 \GeV$)
has emerged. The curve $\mu(A_0)$ becomes flat around
$A_0 \approx 1000 \GeV$ and SAS3 is not suitable to find solutions in
the region where the two solution branches merge;  the two-scale solver, 
however, is able to find more solutions in this case.  Besides this, the 
multiple solutions that we obtain for SAS3 are again the same as found 
for SAS1 in \sct{ssec:lightSUSY}.

In conclusion, scanning over $\mu$ while receiving either $M_{1/2}$ or
$A_0$ as output does not give us any new solutions compared to the
$m_0^2$ output search strategy. We nevertheless have been able to
reproduce the previous results in a consistent manner and also our
expectation of finding two branches of solutions for parameters of
mass dimension one has been fulfilled.  Furthermore, the same singular
structures for small values of $\mu$ have appeared for SAS2 and SAS3.
Note also, that both the two-scale and the semi-analytic solver had to
be used to find all solutions in the considered parameter regions.

\subsection{Is the CMSSM still excluded?}
\label{ssec:CMSSM_excluded}

In this section we investigate the relevance of multiple solutions for globally
fitting the CMSSM to experimental and observational data. As a reference
point we use \citere{Bechtle:2015nua} (``Killing the CMSSM softly''), in which
the program \textsc{Fittino} was used. The idea was to scan over a reasonable
region of the CMSSM input parameter space and to determine the fit point which
is the most compatible with the considered observables. Furthermore, this paper
was the first to derive a consistent $p$-value for the CMSSM from toy experiment.

\subsubsection{The Results of ``Killing the CMSSM Softly''}
\label{sssec: results of killing}
We briefly list the observables which have been used in the analysis of
\citere{Bechtle:2015nua}.  As for the precision observables there are the anomalous
magnetic moment of the muon $a_\mu$, the effective weak mixing-angle
$\sin{\theta_\text{eff}}$, the top quark and $W$ boson masses
as well as the b quark/B meson branching ratios.  Additionally, different
combinations of Higgs observables, like e.g.\ the SM Higgs boson mass and its
decay channels, as observed by the ATLAS/CMS experiments at the LHC, and the
dark matter relic density $\Omega h^2$, as measured by the Planck collaboration,
are incorporated.

The following parameter values were found to give the best accordance between
measured and predicted observables and are hereinafter referred to as
``best-fit point'' parameters:
\begin{equation}
\begin{aligned}
  &t_\beta = 17.7, &&m_0 = 387.4 \GeV, \\
  &M_{1/2} = 918.2 \GeV, &&A_0 = -2002.8 \GeV.
\end{aligned}
\label{eq:best_fit_point}
\end{equation}
It should be noted that these values were determined with the spectrum
generators \texttt{SPheno} 3.2.4 \cite{Porod:2003um,Porod:2011nf} and
\texttt{FeynHiggs} 2.10.1
\cite{Heinemeyer:1998yj,Heinemeyer:1998np,Degrassi:2002fi,Frank:2006yh,Hahn:2013ria,Bahl:2016brp,Bahl:2017aev,Borowka:2014wla,Heinemeyer:2007aq,Hollik:2014bua,Hollik:2015ema},
whereas the following analysis is based on the predictions of \FS\ 2.1.0.

For these input parameters and $\mu > 0$ \footnote{The solution with negative 
$\mu$ is located in an unphysical region where the $CP$-odd Higgs boson becomes tachyonic.} 
\FS finds
\begin{equation}
\begin{aligned}
  &\mu(\ms) = 1505.5 \GeV, \quad M_1(\ms) =  396.2 \GeV, \\ 
  &M_2(\ms) = 725.3 \GeV.
\end{aligned}
\end{equation}
The chargino masses are determined by $M_2$ and $\mu$.  The lightest chargino is 
thus able to escape the LEP bound $m_{\chi^\pm_1} > 94 \GeV$ \cite{Tanabashi:2018oca}.
The lightest supersymmetric particle (LSP) is the lightest neutralino with mass $\sim 400 \GeV$ 
and provides a dark matter candidate.  At the best-fit point the predicted dark matter relic density 
$\Omega h^2$ is in agreement with the experimental observations.

The Standard Model prediction for the anomalous magnetic moment of the muon $a_\mu$ 
deviates from the observed value at a $3.5 \sigma$ level.  To account for this, a successful 
SUSY model is expected to give an additional contribution to $a_\mu$ of the order 
${30 \times 10^{-10}}$ \cite{Davier:2010nc}.  The best-fit point predicts a correction 
${a_\mu^\text{SUSY} \sim 4 \times 10^{-10}}$, which is far too small.

\citere{Bechtle:2015nua} concludes by giving the following $p$-value for the CMSSM:
\begin{equation}
  p = (4.9 \pm 0.7) \%.
\end{equation}

When performing a global fit of a model to known observables, all
possible mathematical solutions to the formulated boundary value
problem need to be taken into account.  The TSS of \texttt{SPheno},
however, which was used in \citere{Bechtle:2015nua}, does not
necessarily find all solutions.  For this reason we study in the
following section whether further solutions to the CMSSM BVP can be
found around the best-fit point with the semi-analytic approach.

\subsubsection{Multiple Solutions around the Best-Fit Point}
\label{sssec: ms around bfp}
\begin{figure}[ht]
  \centering
  \subfigure[]{
  \includegraphics[width = 0.9\columnwidth]{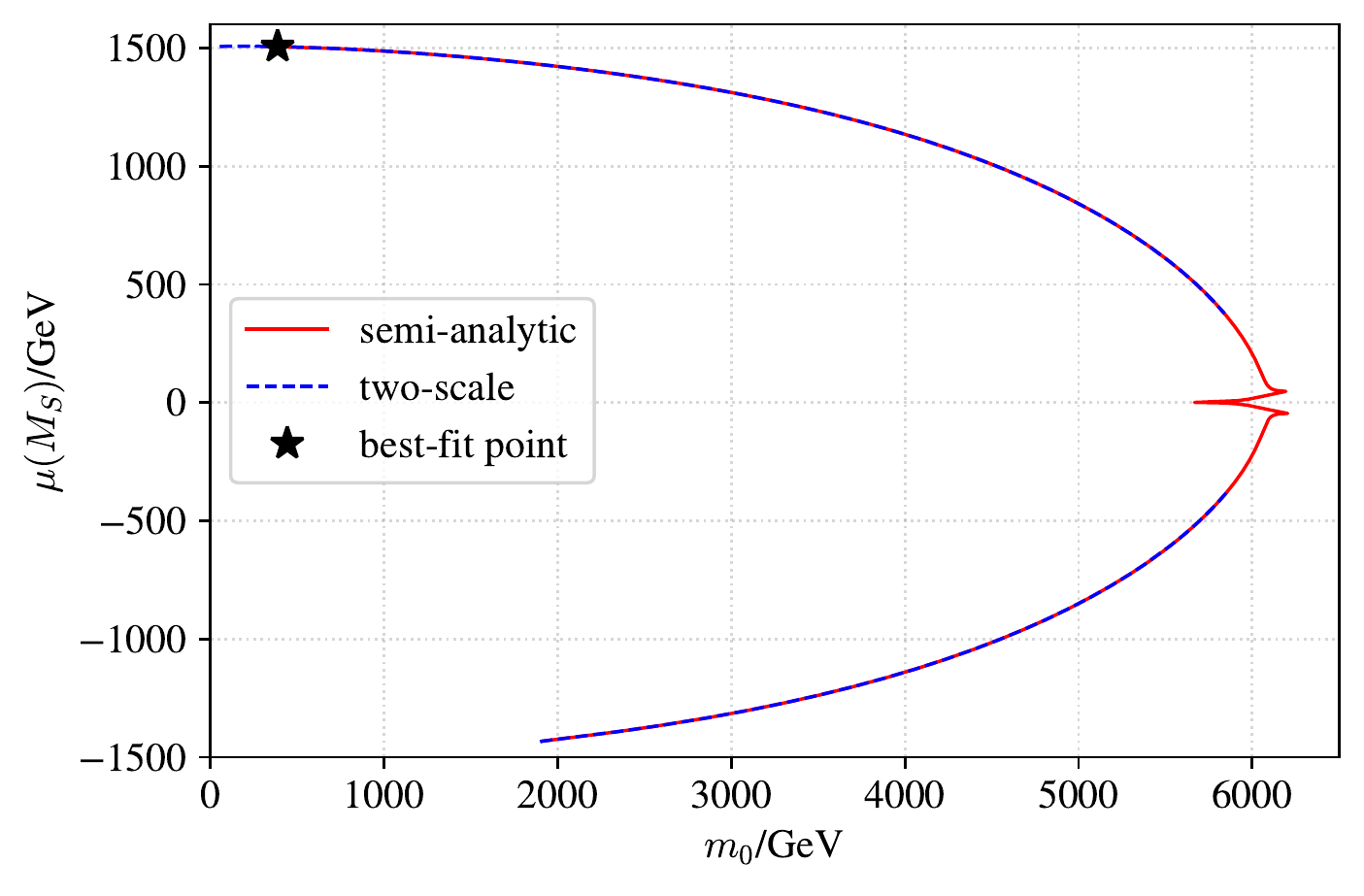}
  \label{fig: bf scan m0}}
  \\
  \subfigure[]{
  \includegraphics[width = 0.9\columnwidth]{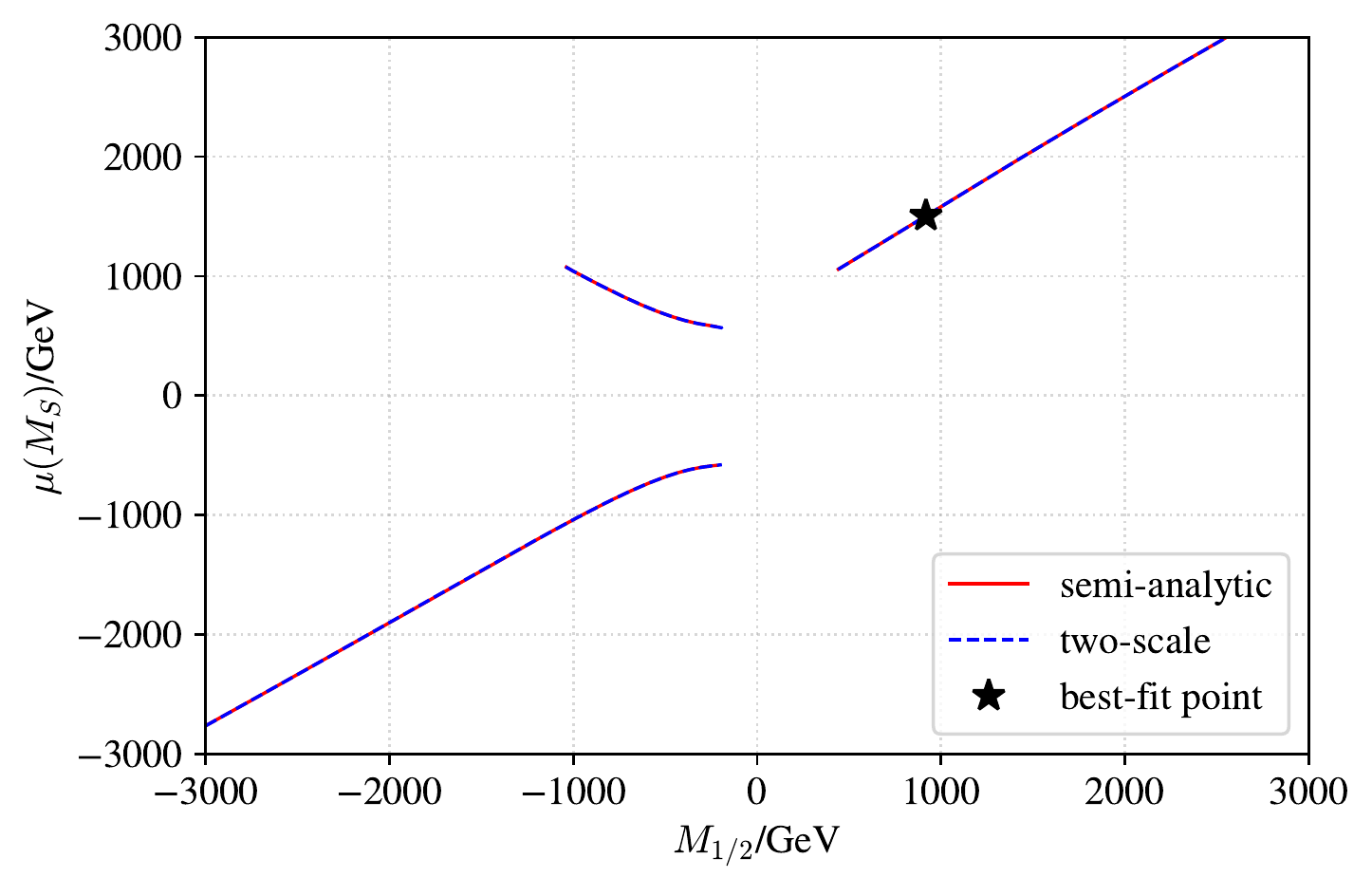}
  \label{fig: bf scan M12}}
  \\
  \subfigure[]{
  \includegraphics[width = 0.9\columnwidth]{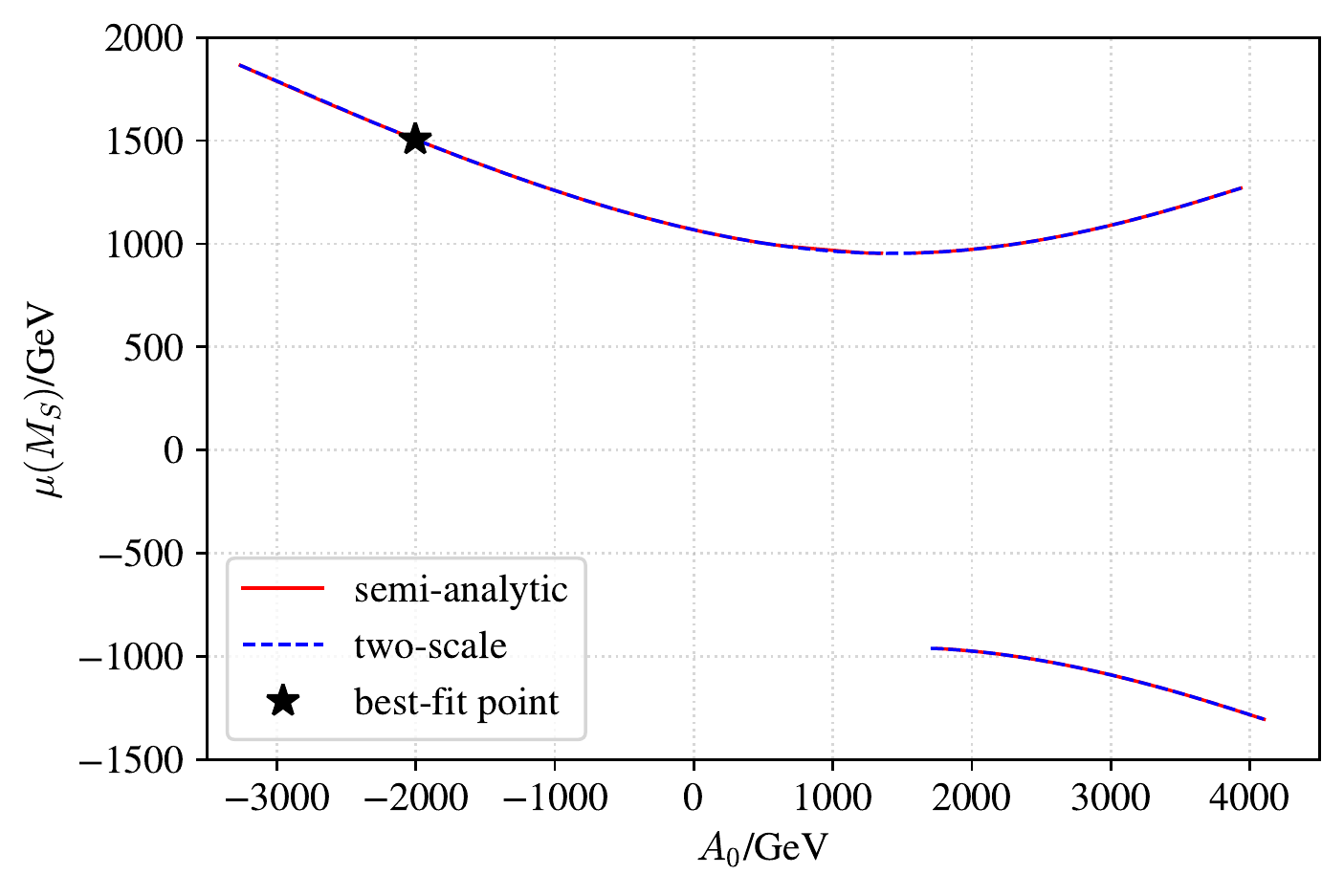}
  \label{fig: bf scan A0}}
  \caption{The CMSSM GUT parameters and $\mu$ are varied around the
  best-fit point from \citere{Bechtle:2015nua} using the two-scale 
  solver and the different semi-analytic solvers (cf. \tab{tab:CMSSM_parameters}).
  In (a) the best-fit point lies  in a region without multiple solutions.
  In (b) and (c) no multiple solutions appear at all---apart from the
  expected ones which differ by $\sign(\mu)$.}
  \label{fig: best fit scan}
\end{figure}

In this section we perform an analysis similar to the one of
\sct{ssec:different_output_pars} for the best-fit point
(cf.\ \eqn{eq:best_fit_point}).  All except one of the free CMSSM GUT input
parameters are set to their best-fit values and the remaining
parameter is varied subsequently with the TSS of $\FS$.  To find
possible multiple solutions we repeat the same procedure but scan over
$\mu(\ms)$ by means of the semi-analytic solvers SAS1--SAS3.  The
results are shown in \figs{fig: best fit scan}.

As for \fig{fig: bf scan m0} we recognize the same overall relation between
$m_0$ and $\mu$ as in \fig{fig:CMSSM_TSS_SAS} and the multiple solutions
around $\mu = 0$ appear as well.  At the lower end of the curve, the multiple
solutions due to non-linear parameter inter-dependencies are non-existent.
There, for instance,
\begin{equation}
  \frac{\text{d} m_{h_u}^2}{\text{d} \mu} \bigg|_{\mu = -1433 \GeV}
  \approx 3112 \GeV,
\end{equation}
and so \eqn{eqn: mHu2_mu} is not fulfilled.

In \figs{fig: bf scan M12} and \ref{fig: bf scan A0} the semi-analytic solver
does not find any additional solutions compared to the two-scale solver.
In the case of \fig{fig: bf scan A0} the curve $\mu(A_0)$ becomes too flat
around $A_0 \approx 1500 \GeV$ to allow an efficient scan over $\mu$
with the SAS3.  For some values of the scan parameters, for instance the
region  where $0 < M_{1/2} < 400 \GeV$ in \fig{fig: bf scan M12}, we do
not find a physical solution due to either tachyonic down-type sleptons,
up-type squarks, or Higgs bosons.

Concluding, we find that the best-fit point is far off from the regions in which
multiple solutions can occur.  One reason is that around the best-fit point all
SUSY particles are heavy enough to escape the experimental constraints,
while our previous analysis has shown that additional solutions tend to occur
in regions of parameter space where at least some superpartners become light.

\section{Multiple solutions in the CNMSSM}
\label{sec:CNMSSM}

In this section we study the occurrence of multiple solutions for a given set of
parameters $\{m_0^2$, $\lambda$, $\kappa\}$ in the CNMSSM.  Exchanging
$m_0^2$ with $\mueff$ allows us to search for multiple solutions in a similar
fashion as we did for the CMSSM.  It has to be noted, however, that we now
also take the GUT parameters $M_{1/2}$ and $A_0$ to be output of our
algorithm and not input as we did in the CMSSM.

\subsection{Study of multiple solutions with the semi-analytic approach}

As described in \sct{sec:CNMSSM_BVP}, the CNMSSM is formulated as a
BVP with the universal GUT parameters $\{m_0^2$, $M_{1/2}$, $A_0\}$.
In order to study multiple solutions in the CNMSSM, we make use of the
semi-analytic equations, which allows us to take the SUSY scale parameters
$\{\lambda$, $\kappa$, $\mueff\}$ as input.  Specifying the dimensionless
quantities $\lambda$ and $\kappa$ yields a good stability of the underlying
solving algorithm.  $\mueff$ can now be used as a scan parameter as was
done for the CMSSM.  As a result, we formulate the CNMSSM BVP in
terms of the input parameters
\begin{equation}
  t_\beta(\mz), \, \lambda(\ms), \, \kappa(\ms), \, \mueff(\ms)
\end{equation}
and obtain $\{m_0^2$, $M_{1/2}$, $A_0\}$ as output.

\begin{figure}[t]
\centering
\subfigure[]{
\includegraphics[width = 0.9\columnwidth]{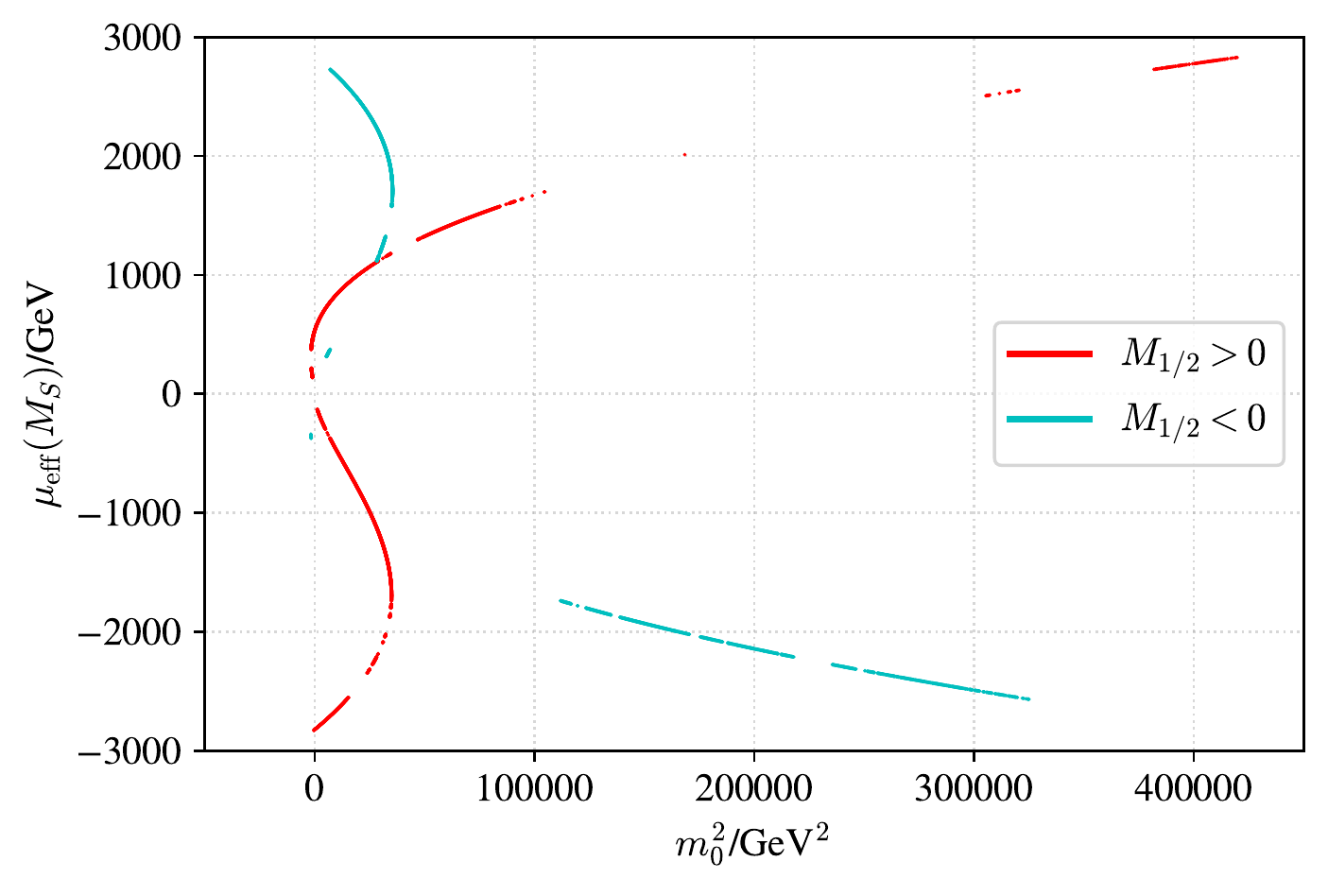}
\label{fig:CNMSSM_scans_m0}
}
\\
\subfigure[]{
\includegraphics[width = 0.9\columnwidth]{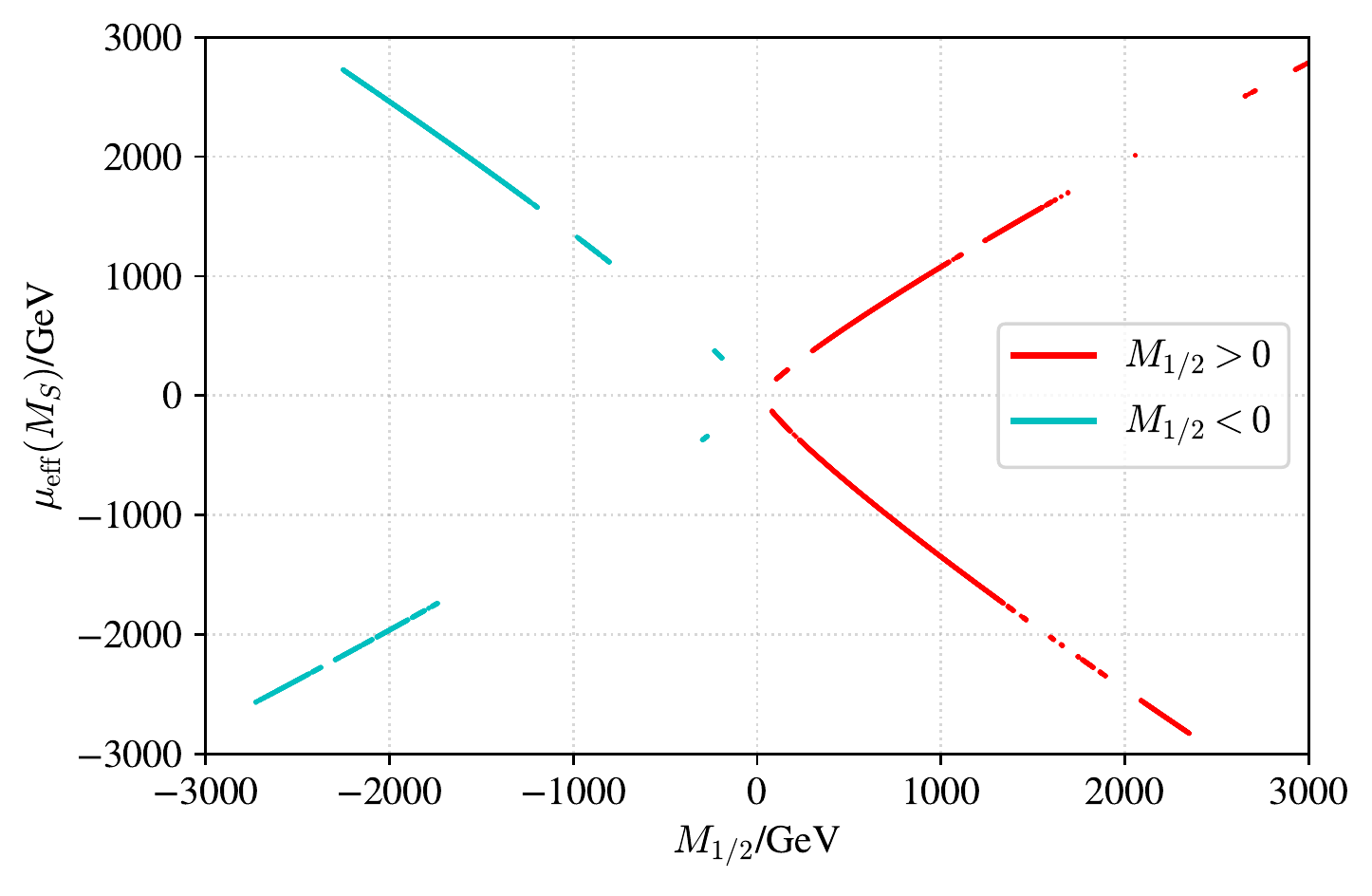}
\label{fig:CNMSSM_scans_M12}
}
\\
\subfigure[]{
\includegraphics[width = 0.9\columnwidth]{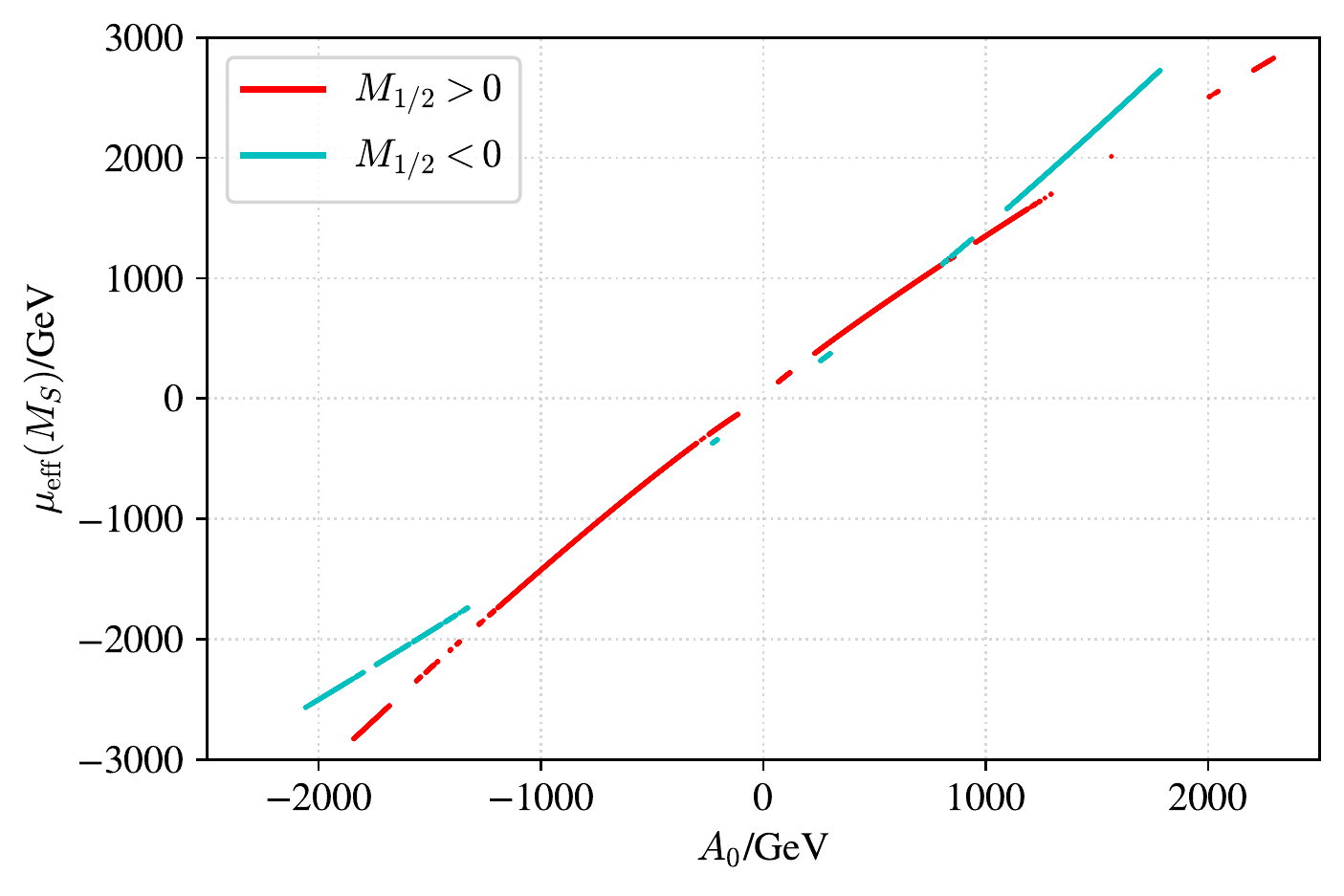}
\label{fig:CNMSSM_scans_A0}
}
\caption{The CNMSSM GUT parameters as functions of $\mueff$. The dimensionless
input parameters are $t_\beta(\mz) = 9$, $\lambda(\ms) = 0.04$, and 
$\kappa(\ms) = -0.013$. The solutions with positive and negative $M_{1/2}$
are plotted in different colors.}
\label{fig:CNMSSM_scans}
\end{figure}
In \figs{fig:CNMSSM_scans} we show a scan over $\mueff(\ms)$ for fixed
values of $t_\beta$, $\lambda$, and $\kappa$.  The output parameters
$\{m_0^2$, $M_{1/2}$, $A_0\}$ are shown on the abscissae.  The curves are
discontinuous due to the existence of two distinct solutions branches, which
differ from each other in their sign of $M_{1/2}$.  In order to distinguish the
branches, solutions with $\sign(M_{1/2}) = \pm 1$ are marked as red and
cyan dots, respectively.  If we were able to pre-select one of the branches,
a scan should yield a continuous relation between the output parameters
and $\mueff$.

In contrast to the CMSSM, where we found multiple solutions, now we do not
find any physical solutions in the parameter region $|\mueff| \lesssim M_{Z,W}$
at all.  The reason for this is that, in our scenario, as $\mueff$ tends to 0,
the dimensionful output parameters become small and tachyons appear in the
particle spectrum, i.e.\ the solutions of the BVP have to be discarded.
Furthermore we find a highly non-linear dependence of $m_0^2$ on $\mueff$
for each $\sign(M_{1/2})$, see \fig{fig:CNMSSM_scans_m0}, which
results in up to three solutions around $m_0^2 \gtrsim 0$.  This
non-linear dependence can be understood as follows:
When determining the three GUT parameters from the three EWSB equations
of the CNMSSM, the strongest restriction to the value of $m_0^2$ comes from
the equation \cite{Ellwanger:2009dp}
\begin{equation}
\begin{aligned}
  m_s^2 &= \lambda \kappa v_d v_u
  + \frac{1}{\sqrt{2}} \lambda A_\lambda \frac{v_d v_u}{v_s}
  - \frac{1}{2} \lambda^2 \left( v_d^2 + v_u^2 \right) \\
  & \phantom{{}={}} \negmedspace - \kappa^2 v_s^2 - \frac{1}{\sqrt{2}}
  \kappa v_s A_\kappa.
\end{aligned}
\end{equation}
For small $\lambda$ and $\kappa$, the parameters $m_s^2$ and $A_\kappa$
are approximately constant between the GUT and the EW scale \cite{Djouadi:2008uj}, i.e.\
$m_s^2(\ms) \approx m_0^2$ and $A_\kappa(\ms) \approx A_0$.
To obtain $\mueff \sim \mathcal{O}(1\TeV)$
while keeping $\lambda$ small, we have to allow for $v_s \gg v_d,v_u$ and can
approximate
\begin{equation}
  m_0^2 \approx - \kappa^2 v_s^2 - \frac{1}{\sqrt{2}} \kappa v_s A_0.
\end{equation}
From \fig{fig:CNMSSM_scans_A0} we can also see $A_0 = a(v_s) v_s$ with
some positive function $a(v_s)$, which depends only weakly on $v_s$, and so
\begin{equation}
  m_0^2 \approx - |\kappa|^2 v_s^2 + \frac{a(v_s)}{\sqrt{2}} |\kappa| v_s^2,
\end{equation}
for $\kappa < 0$. The term with positive sign is only approximately quadratic
in $v_s$ and takes the shape of a slightly tilted parabola. It is the sum of both
terms which causes the function $m_0^2(\mueff)$ to behave in the observed
non-monotonic way.

The relation between $M_{1/2}$ and $\mueff$ is nearly linear, as one
would expect for two parameters of mass dimension one. The two
branches with different $\sign(M_{1/2})$ are related to one another by
an approximate central symmetry, see \fig{fig:CNMSSM_scans_M12}.
The reason for the resulting discontinuous cross-like shape is that, as
explained above, for a given value of $\mueff$ the semi-analytic solver
usually can find a solution for one value of $\sign(M_{1/2})$, but not
simultaneously for the opposite one.
A similar behavior can be found for $A_0(\mueff)$ in
\fig{fig:CNMSSM_scans_A0}, where the semi-analytic solver can find
one solution for fixed $\mueff$ at most. Here, however, the sign of
$A_0$ is fully determined by the sign of $\mueff$ and the output
parameter differs only slightly between the two solution branches.

\subsection{Mass spectra for two different solutions of a single
  CNMSSM parameter point}

\begin{figure}[t]
  \centering
  \subfigure[]{
  \includegraphics[width = 0.9\columnwidth]{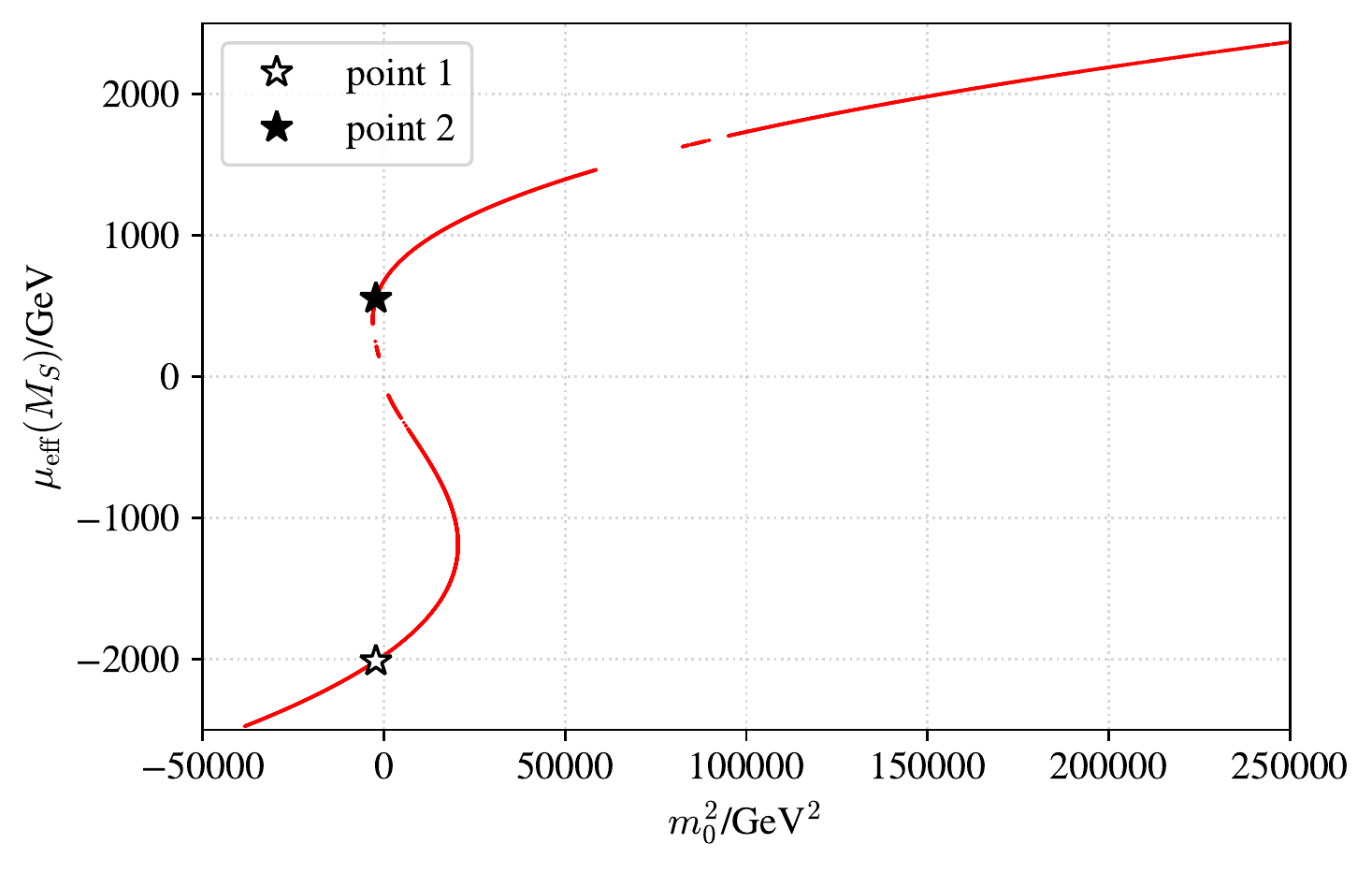}
  \label{fig:CNMSSM_points}}
  \\
  \subfigure[]{
  \includegraphics[width = 0.9\columnwidth]{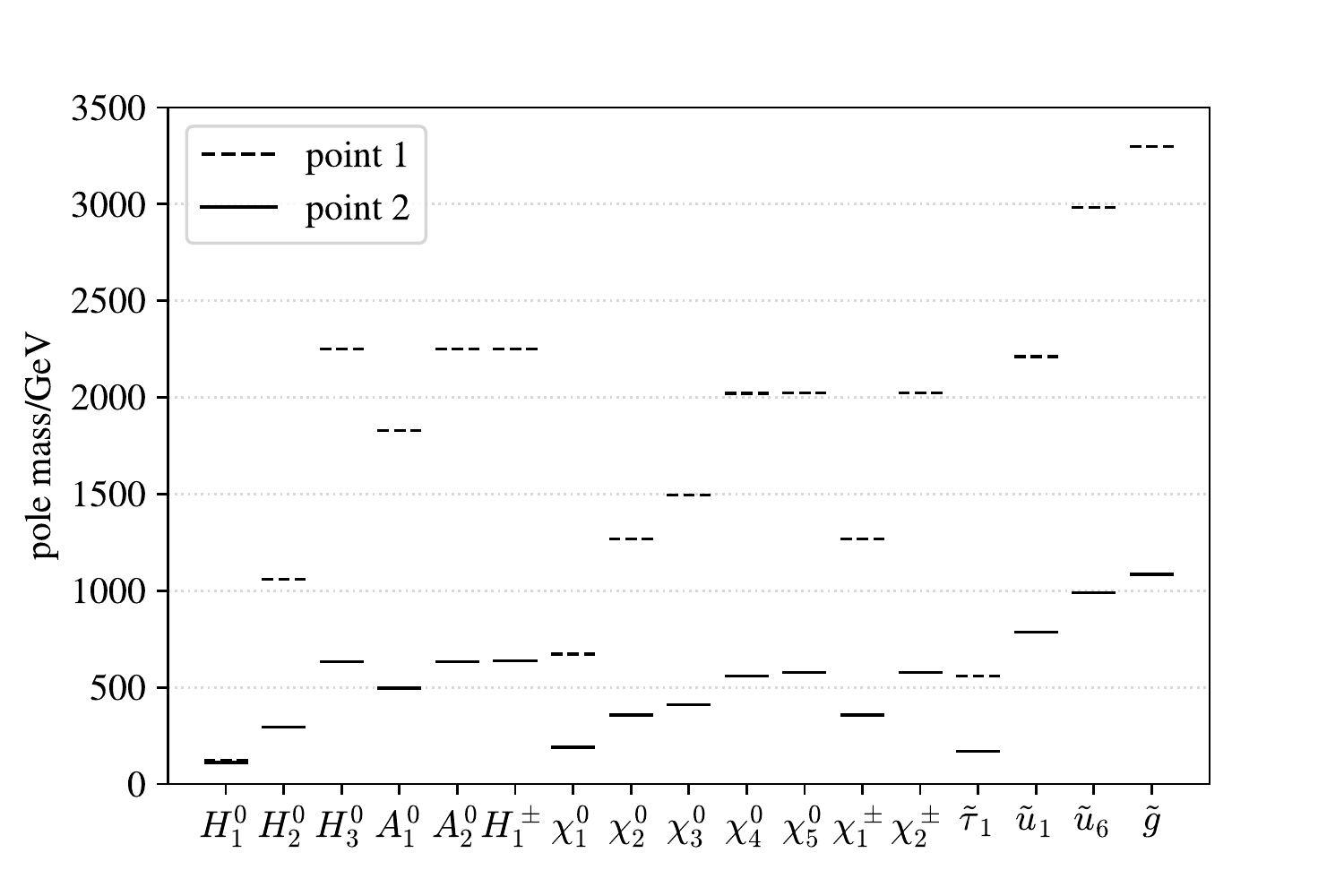}
  \label{fig:CNMSSM_spectra}}
  \caption{Two parameter points fulfilling the CNMSSM boundary condition
  given in \eqn{eqn:CNMSSM_BVP}. The points differ in their values for
  $\mueff$, $M_{1/2}$, and $A_0$ (see \tab{tab:CNMSSM_parameter_points}),
  which results in two vastly distinct pole mass spectra shown in (b).}
  \label{fig: CNMSSM spectra}
\end{figure}

In order to see the importance of multiple solutions in the CNMSSM, we
compare the mass spectra of two different solutions for the parameter point
\begin{equation}
\begin{aligned}
  &m_0^2 = -2200 \GeV^2, && t_\beta(\mz) = 9, \\
  &\lambda(\ms) = 0.035, && \kappa(\ms) = -0.013.
\end{aligned}
\label{eqn:CNMSSM_BVP}
\end{equation}
The two solutions share the same $m_0^2$, but have different
$M_{1/2}$, $A_0$ and $\mueff$, see \tab{tab:CNMSSM_parameter_points}
and \figs{fig: CNMSSM spectra}.
\begin{table}[b]
  \caption{CNMSSM parameter points}
  \label{tab:CNMSSM_parameter_points}
  \begin{tabular}{lrrr}
    \hline\noalign{\smallskip}
    & $\bigwhitestar$ point 1 & $\bigstar$ point 2 \\
    \hline\noalign{\smallskip}
    $\mueff$/GeV & $-2013$ & $553$ \\
    $M_{1/2}$/GeV & $1556$ & $469$ \\
    $A_0$/GeV & $-1495$ & $401$ \\
    \noalign{\smallskip}\hline\noalign{\smallskip}
    $M_h$/GeV & $119.5$ & $110.3$ \\
    $M_{\chi^0_1}$/GeV & 673.3 & 189.7 \\
    $M_{\tilde{\tau}_1}$/GeV & $558.1$ & $169.8$ \\
    \noalign{\smallskip}\hline
  \end{tabular}
\end{table}
As a result, the points have different pole mass spectra as shown
\fig{fig:CNMSSM_spectra}.  By comparing the predicted Higgs boson pole
masses with the experimentally measured value of
$M_h = (125.10 \pm 0.14)\GeV$, point 2 can be excluded, while point 1
may still be taken into consideration.  However, none of the points
can correctly predict the observed Dark Matter relic density because
the lightest stau is the lightest supersymmetric particle (LSP).  This
is in agreement with the analysis of \citere{Djouadi:2008uj}, finding
that the condition $A_0 \sim -M_{1/2}/4$ must be fulfilled in order
for a CNMSSM parameter point to produce the observed dark matter
relic density.

In any case, our analysis shows that the different solutions can have
a significantly different phenomenology and the SAS is a useful tool
to find and study multiple solutions in this model.  However,
since $M_{1/2}$ and $A_0$ are output parameters in our SAS formulation
of the CNMSSM BVP, viability conditions such as the ones given in
\citere{Djouadi:2008uj} cannot be enforced from the start and have to be
tested on the output parameters.

\section{Conclusions}

In \citeres{Allanach:2013yua,Allanach:2013cda,Allanach:2014sea} the
appearance of multiple solutions to the BVP of the CMSSM has been
discovered and studied.  In the present paper we have investigated the
deeper origin of these multiple solutions.
The study was made possible by the semi-analytic BVP solver
implemented in \FS, which allows to exchange input and output
parameters to search for turning points of inverse functions of the
BVP and thus allows the systematic search for multiple solutions.
We could trace their appearance back to two phenomena:
\begin{itemize}
\item Light neutralinos and charginos can lead to singular points in
  the one-loop $W^\pm$ and $Z^0$ self-energies, which translate to
  singular points in the function $m_0^2(\mu)$.  At these points the
  derivative of $m_0^2(\mu)$ can change its sign, which leads to
  multiple branches in the inverse functions $\mu(m_0^2)$.  The
  position of the singular points is given by the light
  neutralino/chargino masses.  The number of singular points depends
  on their couplings to the $W^\pm$ and $Z^0$ bosons and on the
  formulation used to determine the \DR weak mixing angle from
  physical observables.
\item A non-linear inter-dependence between the parameters $m_0^2$ and
  $\mu$ can lead to a minimum of the function $m_0^2(\mu)$, resulting
  in the appearance of multiple branches in the inverse function
  $\mu(m_0^2)$ around that minimum.
\end{itemize}
Furthermore we have answered the question whether the CMSSM is still
excluded in the presence of potential multiple solutions.  We find that around
the CMSSM best-fit point the solution to the BVP is unique and thus
the $p$-value for the CMSSM remains at 4.9\%.

Finally we have investigated the appearance of multiple solutions in
the CNMSSM.  We find that:
\begin{itemize}
\item Multiple solutions around $\mueff \lesssim M_{W,Z}$ tend to not
  occur, because in the limit $\mueff\to 0$ also $m_0^2$ and other
  supersymmetry-breaking parameters vanish or become of the order of
  the electroweak scale, which leads to light or tachyonic scalar
  particles, i.e.\ unphysical solutions of the BVP.
\item For small $m_0^2$ up to three solutions for $\mueff$ can occur
  due to a non-linear inter-dependence between these two parameters,
  imposed by the EWSB equations and the $\beta$ functions.  The
  different solutions may have significantly different physical
  spectra because of different values for $M_{1/2}$, $A_0$ and
  $\mueff$.
\end{itemize}
Concluding, we would like to emphasize that in order to investigate
the validity of a constrained SUSY model all possible solutions to the
BVP must be studied.  In combination with the conventional ``running
and matching'' procedure (TSS), the semi-analytic approach (SAS) is a
useful tool to search for additional solutions.

\begin{acknowledgements}
  We kindly thank Ben Allanach, Peter Athron, Dylan Harries and Werner
  Porod for helpful discussions.
\end{acknowledgements}

\bibliographystyle{spphys}
\bibliography{paper}

\end{document}